\documentclass[preprint,12pt,authoryear]{elsarticle}

\usepackage{amssymb}
\usepackage{amsthm}
\usepackage{amsmath}

\usepackage{color}

\usepackage{comment}

\journal{Discrete Applied Mathematics}

\usepackage{hyperref}
\hypersetup{colorlinks=true,linkcolor=blue,filecolor=magenta,urlcolor=cyan}
\newtheorem{fact}{Fact}
\newtheorem{lemma}[fact]{Lemma}
\newtheorem{proposition}[fact]{Proposition}
\newtheorem{theorem}[fact]{Theorem}
\newtheorem{corolary}[fact]{Corollary}

\newcommand{\alphabet}{\Sigma}
\newcommand{\cjtoAlinha}[1]{\mathcal{A}_{{#1}}}
\newcommand{\cjtoAlinhaExt}[1]{\mathcal{E}_{{#1}}}

\newcommand{\alinhamento}[1]{[{#1}]}
\newcommand{\alinhamentoA}[1]{\left[ {#1} \right]}
\newcommand{\alinhamentoB}[1]{\Big[ {#1} \Big]}

\DeclareMathAlphabet{\mathpzc}{OT1}{pzc}{m}{it}

\newcommand{\fonte}{\mathpzc}
\newcommand{\fonteDois}{\mathbb}
\newcommand{\metricaC}{\fonteDois{M}^{\text{C}}}
\newcommand{\metricaE}{\fonteDois{M}^{\text{E}}}
\newcommand{\metricaA}{\fonteDois{M}^{\text{A}}}
\newcommand{\metricaN}{\fonteDois{M}^{\text{N}}}

\newcommand{\metrica}{\fonte{M}}
\newcommand{\prametrica}{\fonte{Pr}}

\newcommand{\hemimetrica}{\fonte{H}}

\newcommand{\pseudometrica}{\fonte{P}}

\newcommand{\semimetrica}{\fonte{S}}

\newcommand{\quasimetrica}{\fonte{Q}}

\newcommand{\matriz}{\gamma}

\newcommand{\pont}[3]{{#1}_{{#2} \rightarrow {#3}}}
\newcommand{\tamanho}[1]{|{#1}|}

\newcommand{\opt}{\ensuremath\protect\mathrm{opt}}
\newcommand{\cost}{\ensuremath\protect{w}}

\newcommand{\distanciaA}[1]{\opt{\rm A}_{{#1}}}
\newcommand{\distanciaN}[1]{\opt{\rm N}_{{#1}}}
\newcommand{\distanciaE}[1]{\opt{\rm E}_{{#1}}}

\newcommand{\custoA}[1]{v{\rm A}_{{#1}}}
\newcommand{\custoN}[1]{v{\rm N}_{{#1}}}
\newcommand{\custoE}[1]{v{\rm E}_{{#1}}}

\newcommand{\seqVazia}{\varepsilon}
\newcommand{\A}{\texttt{a}}
\newcommand{\B}{\texttt{b}}
\newcommand{\C}{\texttt{c}}
\newcommand{\D}{\texttt{d}}

\newcommand{\espaco}{\mbox{-}}

\newcommand{\yes}{$\star$}
\newcommand{\Maior}{\mathcal{Q}}
\newcommand{\maior}{q}

\newcommand{\abs}[1]{\left\vert#1\right\vert}

\newcommand{\portugues}[1]{}

\newcommand\EA[1]{{\color{blue}(EA: #1)}}

\begin{document}

\begin{frontmatter}



\title{Matrices inducing generalized metric on sequences}


\author[UFMS]{Eloi Araujo}
\ead{francisco.araujo@ufms.br}

\author[UFMS]{Fábio V.~Martinez}\corref{corr}
\ead{fabio.martinez@ufms.br}

\author[UFMS]{Carlos H.~A.~Higa}
\ead{carlos.aguena@ufms.br}

\author[USP]{José Soares}
\ead{jose@ime.usp.br}

\cortext[corr]{Corresponding author.}

\address[UFMS]{Faculdade de Computação, Universidade Federal de Mato Grosso do Sul, Brasil}

\address[USP]{Departamento de Ciência da Computação, Instituto de Matemática e Estatística, Universidade de São Paulo, Brasil}

\begin{abstract}
Sequence comparison is a basic task to capture similarities and 
differences between two or more sequences of symbols, with countless 
applications such as in computational biology. An alignment is a way to compare sequences, where a giving scoring function determines the degree of similarity between them. Many scoring functions are obtained from scoring matrices. However, not all scoring matrices induce scoring functions which are distances, since the scoring function is not necessarily a metric. In this work we establish necessary and sufficient conditions for scoring matrices to induce each one of the properties of a metric in weighted edit distances. For a subset of scoring matrices that induce normalized edit distances, we also characterize each class of scoring matrices inducing normalized edit distances. Furthermore, we define an extended edit distance, which takes into account a set of editing operations that transforms one sequence into another regardless of the existence of a usual corresponding alignment to represent them, describing a criterion to find a sequence of edit operations whose weight is minimum. Similarly, we determine the class of scoring matrices that induces extended edit
distances for each of the properties of a metric.
\end{abstract}


\end{frontmatter}


\section{Introduction} \label{sec:intro}

Sequence comparison is a classical problem in computer science and
has applications in several areas such as computational biology, text
processing, and pattern recognition. A scoring function is a measure to determine the degree of similarity between two sequences and can be defined by edit operations that transform one sequence into another. Typical edit operations are insertion, deletion, and substitution of symbols.

A simple scoring function to compare two given sequences $s$ and $t$ is the Levenshtein distance~\citep{Levenshtein1965}, or \emph{edit distance}, defined as the minimum number of edit operations that transform $s$ into $t$. Given that, the edit distance is a way of quantifying the  dissimilarity of two strings and, as a consequence, it is used in several applications where the data can be represented by strings. For instance, in~\citep{BARTON2015} the edit distance was used in an application related to genome assembly where the authors showed that adding the flexibility of bounding the number of gaps inserted in an alignment strengthens the classical sequence alignment scheme of scoring matrices.

There are some variations of the edit distance that worth mentioning. When each symbol in $s$ must be edited exactly once, one can usually compare $s$ and $t$ by computing the \emph{weighted edit distance}, that is, the minimum weight to transform $s$ into $t$ through a sequence of weighted edit operations. It is used, for instance, in spelling correction systems~\citep{GOSH2016}. Another function is the \emph{normalized edit distance}~\citep{MarzalV1993}, where the length of the two strings is taken into account when computing the distance between them. The result of this function is the minimum average weight of a set of edit operations required to transform $s$ into $t$. Examples of applications of the normalized edit distance are the text reading from street view images~\citep{SUN2019} and software verification. In the latter, runs of a system are represented using words and it is customary to analyze the relationship between the set of words that satisfy a given specification and the set  of words that the system under examination produces~\citep{FISMAN2022}. The \emph{generalized edit distance} proposed by \cite {YujianB2007} is a metric and it is a simple function of the strings lengths; it was applied in a handwritten digit recognition study and showed that it can generally provide similar results to some other normalized edit distances. Another variation is the \emph{contextual normalized edit distance}, where the cost of each edit operation is divided by the length of the string on which the edit operation takes place. It was introduced by \cite{MICO2008} and they showed that this variation is useful for classification purposes.

Considering the weighted edit distance, any set of edit operations that transforms $s$ into $t$, restricted so that each symbol of $s$ must be edited once and only once, can be represented by an alignment. Consider the following alignment:
\[
\left[
  \begin{array}{cccccc}
    \A & \A & \B & \espaco &\A  & \espaco\\
    \A & \espaco & \A & \B & \B & \A
    \end{array}
  \right]\enspace .
\]
It represents the transformation of $s = \A \A \B \A$ into $t = \A \A \B \B \A$ by replacing the first and third symbol with $\A$, the last one with $\B$, deleting the second symbol, and inserting a $\B$ and an $\A$ before and after the last symbol in $s$. Alignments have been a standard way to compare two or more sequences with a myriad of applications in computational biology~\citep{NeedlemanW1970,SW1981,LP1985,Altschul.et.al.1990,LAK1989,Chenna.et.al.2003,NHH2000,KBH1998,ARRM2021}. 

Let $\alphabet$ be an alphabet and $\alphabet_{\espaco} = \alphabet \cup \{ \espaco \}$, where $\espaco$ is a symbol not belonging to $\alphabet$ which represents insertion and deletion operations. A \emph{scoring matrix} $\matriz$ for $\alphabet$ has its rows and columns indexed by elements of~$\alphabet_{\espaco}$ and represents the weighted edit operations. We denote the entry of $\matriz$ in row $\A$ and column $\B$ by $\pont{\matriz}{\A}{\B}$. If $\A, \B \in \alphabet$, then $\pont{\matriz}{\A}{\B}$ is the weight of replacing the symbol $\A$ with $\B$, $\pont{\matriz}{\A}{\espaco}$ is the weight of deleting the symbol $\A$, $\pont{\matriz}{\espaco}{\B}$ is the weight of inserting the symbol $\B$, and $\pont{\gamma}{\espaco}{\espaco}$ is not defined. Notice that the Levenshtein distance is a weighted edit distance when $\pont{\gamma}{\A}{\B} = 0$ for $\A = \B$, and $\pont{\gamma}{\A}{\B} = 1$ otherwise, for each pair of symbols $\A$ and $\B$.

Denote by $\alphabet^{*}$ the set of all finite sequences on $\alphabet$. Many different scoring functions based on edit operations
are induced by scoring matrices. \citet{Pea2013} presented ways to select scoring matrices with practical significance in bioinformatics, to estimate distance and similarity.

For a matrix $\gamma$, we have that $\opt_\gamma: \alphabet^{*} \times \alphabet^{*} \rightarrow \mathbb{R}$ represents a general scoring function \emph{induced} by $\gamma$ such that $\opt_\gamma(s, t)$ is the minimum score of the edit operations transforming sequence $s$ into sequence $t$. The scoring function $\opt_\gamma$ is a metric on sequences if $\opt_\gamma$ satisfies the axioms of reflexivity, non-negativity, positivity, symmetry, and triangle inequality; and we say that $\gamma$ \emph{induces} an $\opt_\gamma$-metric on sequences.
If $\opt_{\gamma}$ satisfies the property $p$, we say that $\gamma$ \emph{induces} $\opt_{\gamma}\text{-}p$ on sequences.
  
The weighted edit distance between sequences $s$ and $t$ is denoted by
$\distanciaA{\gamma}$. Observe that not all scoring matrices induce 
scoring functions that are distances, since the scoring function is not necessarily a metric. \citet{Sellers1974} described a sufficient condition for a weighted edit distance be a metric on $\alphabet^{*}$, 
\textit{i.e.}, proved that a scoring matrix $\matriz$ induces an 
$\distanciaA{\gamma}$-metric on sequences. \citet{AraujoS2006} presented necessary and sufficient conditions for a scoring matrix $\matriz$ induce a weighted edit distance as a metric on $\alphabet^{*}$. For example, the scoring matrix
\[
\begin{array}{c|cccc}
  \gamma & \A & \B & \C & \espaco\\
  \hline
  \A & 0 & 1 & 3 & 1\\
  \B & 1 & 0 & 1 & 1\\
  \C & 4 & 1 & 0 & 1\\
  \espaco & 1 & 1 & 1 & \espaco
\end{array}
\]
induces a weighted edit distance, although $\pont{\matriz}{\A}{\C} \not= \pont{\matriz}{\C}{\A}$, $\pont{\matriz}{\A}{\C} \not\le \pont{\matriz}{\A}{\espaco} + \pont{\matriz}{\espaco}{\C}$, and $\pont{\matriz}{\C}{\A} \not\le \pont{\matriz}{\C}{\espaco} + \pont{\matriz}{\espaco}{\A}$. Moreover, there are many ways to generalize the concept of metric by relaxing the axioms that define the metric space, which gives rise to different generalized metric spaces such as quasimetric, semimetric, and others. In this work we extended investigations from~\citet{AraujoS2006} establishing necessary and
sufficient conditions for scoring matrices $\gamma$ to induce $\distanciaA{\gamma}\text{-}p$ for each axiom $p$ of a metric on sequences. 

\citet{MarzalV1993} set another scoring function based on edit operations and induced by a scoring matrix $\gamma$, called normalized edit distance and denoted by $\distanciaN{\gamma}$. Given sequences $s$ and $t$, considering that each symbol in $s$ must be edited exactly once, then $\distanciaN{\gamma}(s, t)$ is the minimum average weight
of a set of edit operations required to transform $s$ into $t$.
Similarly, not all scoring matrices $\gamma$ induce an $\distanciaN{\gamma}$-metric and \cite{AraujoS2006} characterize the
class of scoring matrices inducing normalized edit distances. In this work we also characterize each class of scoring matrices inducing 
$\distanciaN{\gamma}\text{-}p$ on sequences for some generalized metric spaces $p$. Recently,~\citet{FISMAN2022} proved that the normalized edit distance proposed in \citep{MarzalV1993} is a metric when cost of all edit operations are the same. Given the contributions of our work aforementioned, we highlight that this article is more general in the sense that it covers the case treated in \citet{FISMAN2022}.

Additionally to the contributions mentioned above, we define a third scoring function based on edit operations induced by scoring matrices $\gamma$, called extended edit distance and denoted by $\distanciaE{\gamma}$. This scoring function takes into account a set of edit operations that transforms one sequence $s$ into another sequence $t$, with no restriction to the number of edit operations in each symbol of $s$, \textit{i.e.}, a symbol in $s$ can be edited an arbitrary number of times. We characterize each class of scoring matrices inducing $\distanciaE{\gamma}\text{-}p$, for each axiom $p$ of a metric.

Summarizing, the main contributions of this paper are:
\begin{itemize}
\item We extended investigations from~\citet{AraujoS2006} establishing necessary and sufficient conditions for scoring matrices $\gamma$ to induce $\distanciaA{\gamma}\text{-}p$ for each axiom $p$ of a metric on sequences;
\item We characterize each class of scoring matrices inducing 
$\distanciaN{\gamma}\text{-}p$ on sequences for some generalized metric spaces $p$; and 
\item We define a scoring function based on edit operations induced by scoring matrices $\gamma$, called extended edit distance, denoted by $\distanciaE{\gamma}$; we characterize each class of scoring matrices inducing $\distanciaE{\gamma}\text{-}p$ for each axiom $p$ of a metric.
\end{itemize}

This paper is organized as follows. Section~\ref{sec:preliminares}
provides a brief description of basic concepts and characterizes the classes of matrices that induce metric properties for the aimed scoring functions: weighted edit distance, normalized edit distance and extended edit distance. Sections \ref{sec:regular}, \ref{sec:normalizado}, and \ref{sec:estendido} present the main results for these distances. Section~\ref{sec:conclusion} concludes.

\section{Preliminaries}\label{sec:preliminares}

\subsection{Sequences, alignments, and scoring functions}

An \emph{alphabet} is a finite non-empty set of symbols. A \emph{sequence} over an alphabet $\Sigma$ is a finite ordered list of symbols from $\Sigma$ and the set of all sequences over $\Sigma$ is denoted by $\Sigma^*$.
We denote a sequence $s \in \Sigma^*$ by $s = s(1)s(2) \cdots s(n)$, where $s(i) \in \Sigma$ is the $i$-th symbol of $s$ and $n = \abs{s}$ is the
\emph{length} of $s$. We denote by $\seqVazia$ the sequence with zero symbols, also called the \emph{empty sequence}. If $s$ and $t$ are sequences over $\Sigma$, then $st$ is a sequence that represents the \emph{concatenation} of $s$ and $t$. The concatenation of $n$ copies of $s$ has length $n|s|$ and is denoted by $s^{n}$. 

Let $\alphabet_ {\espaco} = \alphabet \cup \{\espaco\}$, with $\espaco \not\in \alphabet$. The symbol $\espaco$ is a \emph{space} and represents an insertion or a deletion. An \emph{alignment} of sequences $s, t \in \Sigma^*$ is a pair of sequences $\alinhamento{s', t'}$ obtained by inserting spaces into $s$ and $t$ such that $\tamanho{s'} = \tamanho{t'}$ and there is no $j$ such that $s'(j) = t'(j) = \espaco$. An alignment can be seen as a sequence of edit operations that transforms $s$ into $t$ when each symbol in $s$ must be edited precisely once. We say that the pair of symbols $[s'(j), t'(j)]$ is \emph{aligned} in the $j$-th column of $ \alinhamento {s', t'}$. We also say that $\tamanho{\alinhamento{s', t'}} = \tamanho{s'} = \tamanho{t'}$ is the \emph{length} of alignment $\alinhamento{s', t'}$. We denote the set of all alignments of $s, t$ by $ \cjtoAlinha{s, t}$.
An alignment $\alinhamento{s', t'}$ can be seen by placing $s'$ above $t'$ as follows 
\[
\alinhamentoA{
\begin{array}{cccccccccc}
\A & \C & \espaco & \C & \B & \espaco & \B & \B & \B & \espaco\\
\C & \espaco & \A & \espaco & \A & \C & \espaco & \espaco & \C & \B
\end{array}
} \text{and} 
\alinhamentoA{
\begin{array}{ccccccccccc}
\espaco & \espaco & \espaco & \espaco & 
\A & \C & \C & \B & \B & \B & \B \\
\C & \A & \A & \C & \C & \B & 
\espaco & \espaco & \espaco & \espaco & \espaco
\end{array}
}
\]
where two different alignments of sequences $\A \C \C \B \B \B \B, \C \A \A \C \C \B$ are represented: the alignment $\alinhamento{\A \C \espaco \C \B \espaco \B \B \B \espaco, \C \espaco \A \espaco \A \C \espaco \espaco \C \B}$ at the left side and the alignment $\alinhamento{\espaco \espaco \espaco \espaco \A \C \C \B \B \B \B, \C \A \A \C \C \B \espaco \espaco \espaco \espaco \espaco}$ at the right.


Given a scoring matrix $\matriz$, we define the functions $\custoA{\matriz}$ and $\custoN{\matriz}$ for each alignment $\alinhamento{s', t'}$:  if $s = t = \seqVazia$, $\custoA{\matriz} \alinhamento{s', t'} = \custoN{\matriz} \alinhamento{s', t'} = 0$; otherwise,
\[
\custoA{\matriz} \alinhamento{s', t'} = 
\sum_{j = 1}^{\tamanho{\alinhamento{s', t'}}} 
\pont{\matriz}{s'(j)}{t'(j)} 
\quad \text{and} \quad
\custoN{\matriz} \alinhamento{s', t'} = 
\frac{\sum_{j = 1}^{\tamanho{\alinhamento{s', t'}}} 
\pont{\matriz}{s'(j)}{t'(j)}}{\tamanho{\alinhamento{s', t'}}}\,.
\]
We say that $\custoA{\matriz}[A]$ is the \emph{score} and $\custoN{\matriz}[A]$ is the \emph{normalized score} of alignment $A$.

An \emph{extended alignment} of the sequences $s, t$ is shown as $\alinhamento{c_{1}, \ldots, c_{n}}$, where each $c_{j}$ is a finite sequence with $m_{j} > 0$ symbols in $\alphabet_{\espaco}$, at least one symbol is different from ``$\espaco$'' but no two consecutive symbols are equal to ``$\espaco$'', and $\alinhamento{c_{1}(1) \ldots c_{n}(1), c_{1}(m_{1}) \ldots c_{n}(m_{n})}$ is an alignment of $s, t$. For the sake of simplicity, we say ``alignment'' instead of ``extended alignment'' when it is clear from the context. We say that $c_{j}$ is the $j$-th \emph{column} and $n = \tamanho{\alinhamento{c_{1}, \ldots, c_{n}}}$ is the \emph{length} of the alignment $\alinhamento{c_{1}, \ldots, c_{n}}$. 

Thus, each sequence $c_j$ is represented by a column and it is written from the top to the bottom. For example, $A = \alinhamento{\A, \A\B\C\D, \B\espaco, \espaco\D, \C\A, \espaco\D\A\B\C, \espaco\A\B\C, \D\A\B}$ is an extended alignment of $\A\A\B\C\D, \A\D\D\A\C\C\B$ and it can be represented as
\[
A = \alinhamentoA{
\begin{array}{c}
\A
\end{array} 
\begin{array}{c}
\A \\ \B \\ \C \\ \D
\end{array} 
\begin{array}{c}
\B\\ \espaco
\end{array} 
\begin{array}{c}
\espaco\\ \D
\end{array} 
\begin{array}{c}
\C \\ \A
\end{array} 
\begin{array}{c}
\espaco\\ \D\\ \A\\ \B\\ \C
\end{array} 
\begin{array}{c}
\espaco\\ \A\\ \B\\ \C
\end{array} 
\begin{array}{c}
\D\\ \A \\ \B
\end{array}}.
\]

Notice that, since $\alinhamento{c_{1}(1) \ldots c_{n}(1), c_{1}(m_{1}) \ldots c_{n}(m_{n})}$ is an alignment of $s, t$, we have $c_j(1) \not= \espaco$ or $c_j(m_j) \not= \espaco$ for each $j$.

The \emph{weight} of $c_{j}$ is given by $\custoE{\matriz} [c_{j}] = \sum_{i=1}^{m_{j}-1} \pont{\matriz}{c_{j}(i)}{c_{j}(i+1)}$ and the \emph{score} of the extended alignment $A = \alinhamento{c_{1}, \ldots, c_{n}}$ is
\[
\custoE{\matriz}[A] = \sum_{j} \custoE{\matriz} [c_{j}]\,.
\]

Considering the alignment $A = \alinhamento{\A, \A\B\C\D, \B\espaco, \espaco\D, \C\A, \espaco\D\A\B\C, \espaco\A\B\C, \D\A\B}$ above and that $\pont{\matriz}{\A}{\B} = 0$ if $\A = \B$ and $\pont{\matriz}{\A}{\B} = 1$ if $\A \neq \B$, its score is
\begin{align*}
\custoE{\matriz}[A] &=  \sum_{j} \custoE{\matriz} [c_{j}]\\
&= \custoE{\matriz}[\A] + \custoE{\matriz}[\A\B\C\D]  + \custoE{\matriz}[\B\espaco] + \custoE{\matriz}[\espaco\D]  + \\ 
& \phantom{~=} \custoE{\matriz}[\C\A] + \custoE{\matriz}[\espaco\D\A\B\C]  + \custoE{\matriz}[\espaco\A\B\C]  + \custoE{\matriz}[\D\A\B]\\
&= (0) + (1+1+1) + (1) + (1) + \\
&\phantom{~=} (1) + (1+ 1+ 1 + 1) + (1 + 1 + 1) + (1 + 1) = 15\,.\\
\end{align*}

We denote the set of all extended alignments of $s, t$ by $\cjtoAlinhaExt{s, t}$.

\medskip

We also define the functions $\distanciaA{\matriz},
\distanciaN{\matriz}$ and $\distanciaE{\matriz}$ as follows:
\begin{align*}
\distanciaA{\matriz}(s, t) &= \min_{A \in \cjtoAlinha{s, t}}
\custoA{\matriz}[A]\,, \\
\distanciaN{\matriz}(s, t) &= \min_{A \in \cjtoAlinha{s, t}}
\custoN{\matriz}[A]\,, \\
\distanciaE{\matriz}(s, t) &= \min_{A \in \cjtoAlinhaExt{s, t}}
\custoE{\matriz}[A]\,.
\end{align*}

If $A$ is an alignment and $\custoA{\matriz}[A] =
\distanciaA{\matriz}(s, t)$ or $\custoN{\matriz}[A] =
\distanciaN{\matriz}(s, t)$, we say that $A$ is an \emph{A-optimal} or
\emph{N-optimal} alignment for $\matriz$, respectively. Similarly, if
$A$ is an extended alignment and $\custoE{\matriz}[A] =
\distanciaE{\matriz}(s, t)$, we say that $A$ is an \emph{E-optimal}
alignment for $\matriz$.

\subsection{Weighted digraphs of scoring matrices}

We can represent a scoring matrix as a weighted digraph. Thus, the
weighted digraph for the scoring matrix $\matriz$ is
\[
D(\matriz) = (\alphabet_{\espaco}, ( \alphabet_{\espaco} \times
\alphabet_{\espaco}) \setminus \{ (\espaco, \espaco) \}, \matriz)\,,
\]
where the weight of arc $\A \rightarrow \B$ is $\pont{\matriz}{\A}{\B}$.
A \emph{walk} from vertex $\A_0$ to $\A_n$ in
$D(\gamma)$ is a sequence $W = \A_0, \A_1, \ldots, \A_n$, $\A_i \in \Sigma_{\espaco}$
such that 
$\A_i = \espaco$ implies $\A_{i-1} \not= \espaco$ for each $i = 1, \ldots, n$, and its \emph{weight} is
$\cost(W) = \sum_{i=1}^n \pont{\gamma}{\A_{i-1}}{\A_i}$.
The walk $W$ is also called
\emph{cycle} if $\A_0 = \A_n$ and in this case we can assume that 
$\A_0 = \A_n \not= \espaco$.
If $\cost(W)$ is the minimum weight of any walk from $\A_0$ to $\A_n$, we say that $W$ is an
\emph{optimal walk} from $\A_0$ to $\A_n$.
The weight of an optimal walk from $\A_0$ to $\A_n$ is denoted by $d_{\gamma}(\A_0, \A_n)$
or simply $d (\A_0, \A_n)$ when $\gamma$ is clear in the context.

\subsection{Properties of metric functions}

For a given set $S$, we say that a function $f: S \times S \rightarrow
\mathbb{R}$ is a \emph{metric} on $S$ if $f$ satisfies the following
properties for each $x, y, z \in S$:
\begin{enumerate}
\item[(\textit{i})] $f(x, x) = 0$ (\emph{reflexivity})\,,
\item[(\textit{ii})] $f(x, y) \ge 0$ (\emph{non-negativity})\,,
\item[(\textit{iii})] $f(x, y) > 0$ if $x \not= y$ (\emph{positivity})\,,
\item[(\textit{iv})] $f(x, y) = f(y, x)$ (\emph{symmetry})\,, and
\item[(\textit{v})] $f(x, z) \le f(x, y) + f(y, z)$ (\emph{triangle
  inequality})\,.
\end{enumerate}
Observe that (\textit{ii}) is a direct consequence of (\textit{i}) and (\textit{iii}). Additionally, when the set $S$ is clear from the context, we simply say that $f$ is a metric.

In order to obtain more general functions than metrics, we can relax
the aforementioned properties. For example, for a given set
$S$, we say that $f$ is a \emph{premetric} if $f(x, x) = 0$ and $f(x,
y) \ge 0$ for each ${x, y \in S}$~\citep{Topologia90}.

\begin{figure}[hbt]
  \begin{center}
    \includegraphics[scale=.5]{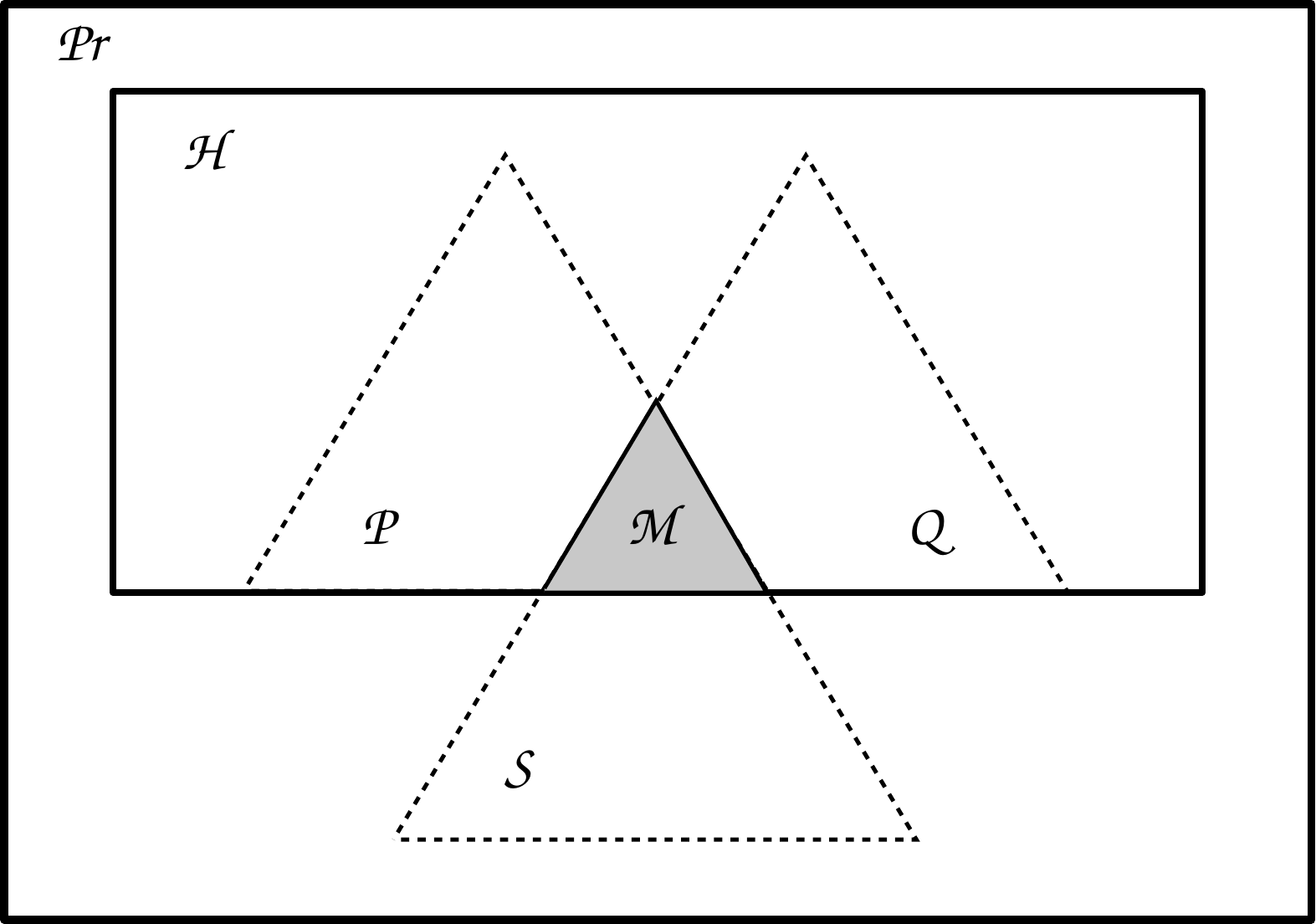}
    \caption{Relationships between functions. Shaded area represents
      the set of metric functions, which is the intersection of all
      the sets.}
    \label{figura-espacos}
  \end{center}
\end{figure}

The table below shows the properties of the following functions:
\emph{premetric} ($\prametrica$), \emph{semimetric}
($\semimetrica$)~\citep{Wilson31-semi}, \emph{hemimetric} or
\emph{pseudoquasimetric} ($\hemimetrica$)~\citep{pseudoQuasi1968},
\emph{pseudometric} ($\pseudometrica$)~\citep{pseudo-1970}, \emph{quasimetric}
($\quasimetrica$)~\citep{Wilson31-quasi} and \emph{metric}
($\metrica$)~\citep{pseudo-1970}. Figure~\ref{figura-espacos} shows
the relationships between these functions.

\begin{center}
\begin{tabular}{lccccccc}
property $\backslash$ function & $\prametrica$ & $\semimetrica$ &
$\hemimetrica$ & $\pseudometrica$ & $\quasimetrica$ & $\metrica$
\\ \hline $f(x, x) = 0, f(x, y) \ge 0$ & yes & yes & yes & yes & yes &
yes \\ $f(x, y) > 0$ for $x \not= y$ & & yes & & & yes & yes \\ $f(x,
y) = f(y, x)$ & & yes & & yes & & yes \\ $f(x, y) \le f(x, z) + f(z,
y)$ & & & yes & yes & yes & yes
\end{tabular}
\end{center}

If a function $\distanciaA{\matriz}, \distanciaN{\matriz}$ or
$\distanciaE{\matriz}$ has a property $p$, we say that the \emph{scoring matrix $\matriz$
induces $\distanciaA{\matriz}\text{-}p, \distanciaN{\matriz}\text{-}p$, or
$\distanciaE{\matriz}\text{-}p$ on sequences}.

A standard class of scoring matrices is $\metricaC$, which has the following properties.
For $\A, \B, \C, \in \alphabet_ {\espaco}$,
\begin{enumerate}
\item[(\textit{i})] $\pont{\matriz}{\A}{\B} > 0$ if $\A \not= \B$,
  and $\pont{\matriz}{\A}{\B} = 0$ if $\A = \B$\,,
\item[(\textit{ii})] $\pont{\matriz}{\A}{\B} =
  \pont{\matriz}{\B}{\A}$\,, and
\item[(\textit{iii})] $\pont{\matriz}{\A}{\C} \le
  \pont{\matriz}{\A}{\B} + \pont{\matriz}{\A}{\C}$\,.
\end{enumerate}
\citet{Sellers1974} showed that scoring matrices in $\metricaC$ induce
an $\distanciaA{\matriz}$-metric on sequences.
The class of scoring matrices $\metricaA$ is such that, for each $\A, \B, \C
\in \alphabet$,
\begin{enumerate}
\item[(\textit{i})] $\pont{\matriz}{\A}{\espaco} =
  \pont{\matriz}{\espaco}{\A} > 0$\,,
\item[(\textit{ii})] $\pont{\matriz}{\A}{\B} > 0$ if $\A \not= \B$,
  and $\pont{\matriz}{\A}{\B} = 0$ if $\A = \B$\,,
\item[(\textit{iii})] if $\pont{\matriz}{\A}{\B} <
  \pont{\matriz}{\A}{\espaco} + \pont {\matriz}{\espaco}{\B}$, then
  $\pont{\matriz}{\A}{\B} = \pont{\matriz}{\B}{\A}$\,,
\item[(\textit{iv})] $\pont{\matriz}{\A}{\espaco} \le
  \pont{\matriz}{\A}{\B} + \pont{\matriz}{\B}{\espaco}$\,, and
\item[(\textit{v})] $\min \{ \pont{\matriz}{\A}{\C},
  \pont{\matriz}{\A}{\espaco} + \pont {\matriz}{\espaco}{\C} \} \le
  \pont{\matriz}{\A}{\B} + \pont{\matriz}{\B}{\C}$\,,
\end{enumerate}
and we show that they contain precisely all
the matrices that induce $\distanciaA{\matriz}$-metric on sequences.
Moreover, \citet{MarzalV1993} mentioned that not every matrix in
$\metricaC$ induces $\distanciaN{\matriz}$-metric on sequences. They showed,
for example, the matrix
\[
\begin{array}{c|ccc}
\gamma & \A & \B & \espaco \\ \hline
\A & 0 & 5 & 5 \\
\B & 5 & 0 & 1 \\
\espaco & 5 & 1
\end{array}
\]
that belongs to $\metricaC$ but 
does not induce $\distanciaN{\matriz}$-triangle inequality since
\begin{align*}
\distanciaN{\matriz} (\A, \B) &= 
\min \left\{ \custoN{\matriz} \alinhamentoB{
\begin{array}{cc}
\A & \espaco\\
\espaco & \B
\end{array}
} ,\; 
\custoN{\matriz} \alinhamentoB{ 
\begin{array}{cc}
\espaco & \A \\
\B & \espaco 
\end{array}
} ,\; 
\custoN{\matriz} \alinhamentoB{ 
\begin{array}{c}
\A\\
\B
\end{array}
}  
\right\} = 3 = \frac{18}{6}\\
&> \frac{17}{6} = \frac{1}{2} + \frac{7}{3} = \custoN{\matriz}  \alinhamentoB{ 
\begin{array}{cc}
\A & \espaco\\
\A & \B
\end{array}
}  + 
\custoN{\matriz} \alinhamentoB{ 
\begin{array}{ccc}
\A & \B & \espaco\\
\espaco & \espaco & \B
\end{array}
} \\
&\ge 
\distanciaN{\matriz}(\A, \A \B) + \distanciaN{\matriz}(\A\B, \B)\,.
\end{align*}
\citet{YujianB2007} pointed out that it was an
open question whether a scoring matrix $\matriz$ induces $\distanciaN{\matriz}$-metric. Nevertheless, we show that matrices in $\metricaN$ induce
$\distanciaN{\matriz}$-metric on sequences. The class of matrices $\metricaN$ is such that
\begin{enumerate}
\item[(\textit{i})] $\metricaN \subseteq \metricaA$\,, and
\item[(\textit{ii})] $\pont{\matriz}{\A}{\espaco} \le
  2\,\pont{\matriz}{\B}{\espaco}$ for each $\A, \B \in \alphabet$\,.
\end{enumerate}  
Discussion above shows that $\metricaC \not\subseteq \metricaN$ and 
we can easily check that scoring matrices $\gamma_1 \in \metricaN \setminus \metricaC$
and $\gamma_2 \in \metricaN \cap \metricaC$, where
\[
\begin{array}{c|ccc}
\gamma_1 & \A & \B & \espaco \\ \hline
\A & 0 & 4 & 1 \\
\B & 3 & 0 & 1 \\
\espaco & 1 & 1
\end{array}
\hspace{1cm}
\mbox{and}
\hspace{1cm}
\begin{array}{c|ccc}
\gamma_2 & \A & \B & \espaco \\ \hline
\A & 0 & 1 & 1 \\
\B & 1 & 0 & 1 \\
\espaco & 1 & 1
\end{array}
\]
which implies that $\metricaN \not\subseteq \metricaC$
and $\metricaC \cap \metricaN \not= \emptyset$.

A third class of scoring matrices $\gamma$ we study is $\metricaE$:
\begin{enumerate}
\item[(\textit{i})] $\pont{\matriz}{\A}{\A} \ge 0$ for each $\A \in
  \alphabet$\,,
\item[(\textit{ii})] $\pont{\matriz}{\A}{\B},
  \pont{\matriz}{\A}{\espaco}, \pont{\matriz}{\espaco}{\A} > 0$ for
  each $\A \not= \B$, $\A, \B \in \alphabet$\,, and
\item[(\textit{iii})] $d_{\matriz}(\A, \B) = d_{\matriz}(\B, \A)$ for each $\A,
  \B \in \Sigma_{\espaco}$\,.
\end{enumerate}

\begin{fact}\label{fato-preliminares}\rm
$\metricaA \subseteq \metricaE$.
\end{fact}

\begin{proof}
  Let $\matriz \in \metricaA$.
  It follows that $\pont{\matriz}{\A}{\espaco} =
  \pont{\matriz}{\espaco}{\A} > 0$, $\pont{\matriz}{\A}{\B} > 0$ if $\A \not= \B$,
  $\pont{\matriz}{\A}{\A} = 0$, 
  $\pont{\matriz}{\A}{\B} = \pont{\matriz}{\B}{\A}$ if $\pont{\matriz}{\A}{\B} <
  \pont{\matriz}{\A}{\espaco} + \pont {\matriz}{\espaco}{\B}$,
  $\pont{\matriz}{\A}{\espaco} \le
  \pont{\matriz}{\A}{\B} + \pont{\matriz}{\B}{\espaco}$, and 
  $\min \{ \pont{\matriz}{\A}{\C},
  \pont{\matriz}{\A}{\espaco} + \pont {\matriz}{\espaco}{\C} \} \le
  \pont{\matriz}{\A}{\B} + \pont{\matriz}{\B}{\C}$ for $\A, \B, \C \in \Sigma$.

  Since $\pont{\matriz}{\A}{\A} = 0$, $\pont{\matriz}{\A}{\espaco} =
  \pont{\matriz}{\espaco}{\A} > 0$, and $\pont{\matriz}{\A}{\B} > 0$ if $\A \not= \B$,
  we have that $\gamma$ satisfies properties
  (\textit{i}) and (\textit{ii}) of $\metricaE$.

Let $W = u_{0}, u_{1}, \ldots, u_{n}$ be an optimal walk
from $\A = u_0$ to $\B = u_n$ in
$D(\matriz)$, and $W_i = u_{i-1}, u_i$. We construct a walk $\overline{W}_{i}$ from $u_{i}$ to
$u_{i-1}$ for each $i = 1, \ldots, n$, in accordance with the cases
below.
\begin{description}
\item[Case 1:] $u_{i-1} = \espaco$ or $u_{i} = \espaco$. Put
  $\overline{W}_{i} = u_{i}, u_{i-1}$.
  Since
  $\pont{\gamma}{u_i}{\espaco} = \pont{\gamma}{\espaco}{u_i}$ and
  $\pont{\gamma}{u_{i-1}}{\espaco} = \pont{\gamma}{\espaco}{u_{i-1}}$,
  it follows that
  $\cost(\overline{W}_{i}) = \pont{\matriz}{u_{i}}{u_{i-1}} =
  \pont{\matriz}{u_{i-1}}{u_{i}} = \cost(W_{i})$\,;
\item[Case 2:] $u_{i-1} \not= \espaco$, $u_{i} \not= \espaco$ and
  $\pont{\matriz}{u_{i-1}}{u_{i}} < \pont{\matriz}{u_{i-1}}{\espaco} +
  \pont{\matriz}{\espaco}{u_{i}}$. Put $\overline{W}_{i} =
  u_{i}, u_{i-1}$.
  Since $\pont{\matriz}{u_{i-1}}{u_{i}} < \pont{\matriz}{u_{i-1}}{\espaco} +
  \pont{\matriz}{\espaco}{u_{i}}$, 
  it follows that $\cost(\overline{W}_{i}) =
  \pont{\matriz}{u_{i}}{u_{i-1}} = \pont{\matriz}{u_{i-1}}{u_{i}} = \cost(W_{i})$\,;
\item[Case 3:] $u_{i-1} \not= \espaco$, $u_{i} \not= \espaco$ and
  $\pont{\matriz}{u_{i-1}}{u_{i}} = \pont{\matriz}{u_{i-1}}{\espaco} +
  \pont{\matriz}{\espaco}{u_{i}}$. Put $\overline{W}_{i} = u_{i}, \espaco, u_{i-1}$.
  Since
  $\pont{\gamma}{u_i}{\espaco} = \pont{\gamma}{\espaco}{u_i}$ and
  $\pont{\gamma}{u_{i-1}}{\espaco} = \pont{\gamma}{\espaco}{u_{i-1}}$,
  it follows that
  $\cost(\overline{W}_i) = \pont{\matriz}{u_{i}}{\espaco} +
  \pont{\matriz}{\espaco}{u_{i-1}} = \pont{\matriz}{u_{i-1}}{\espaco}
  + \pont{\matriz}{\espaco}{u_{i}} =
  \pont{\matriz}{u_{i-1}}{u_{i}} = \cost(W_{i})$\,.
\end{description}

Let $\overline{W}$ be a walk formed by concatenating 
$\overline{W}_{n}, \overline{W}_{n-1}, \ldots, \overline{W}_{1}$.
From the above observations, it follows that
\begin{eqnarray*}
d (u_{0}, u_{n}) = \cost(W) & = &
\sum_{i=1}^{n} \cost(W_{i})\\ & = & 
\sum_{i=1}^{n} \cost(\overline{W}_{i}) =
\cost(\overline{W}) \ge d(u_{n},u_{0})\,.
\end{eqnarray*}
Using similar reasoning, we have $d(u_{n},u_{0}) \ge 
d(u_{0}, u_{n})$.
It follows that 
$d(u_{n},u_{0}) = d(u_{0}, u_{n})$, which implies that $\gamma$ satisfies 
  property (\textit{iii}) of $\metricaE$.

Since $\gamma$ satisfies properties (\textit{i}), (\textit{ii}), and (\textit{iii}),
we have that $\gamma \in \metricaC$.
\end{proof}

\begin{figure}[hbt]
  \begin{center}
    \includegraphics[scale=.6]{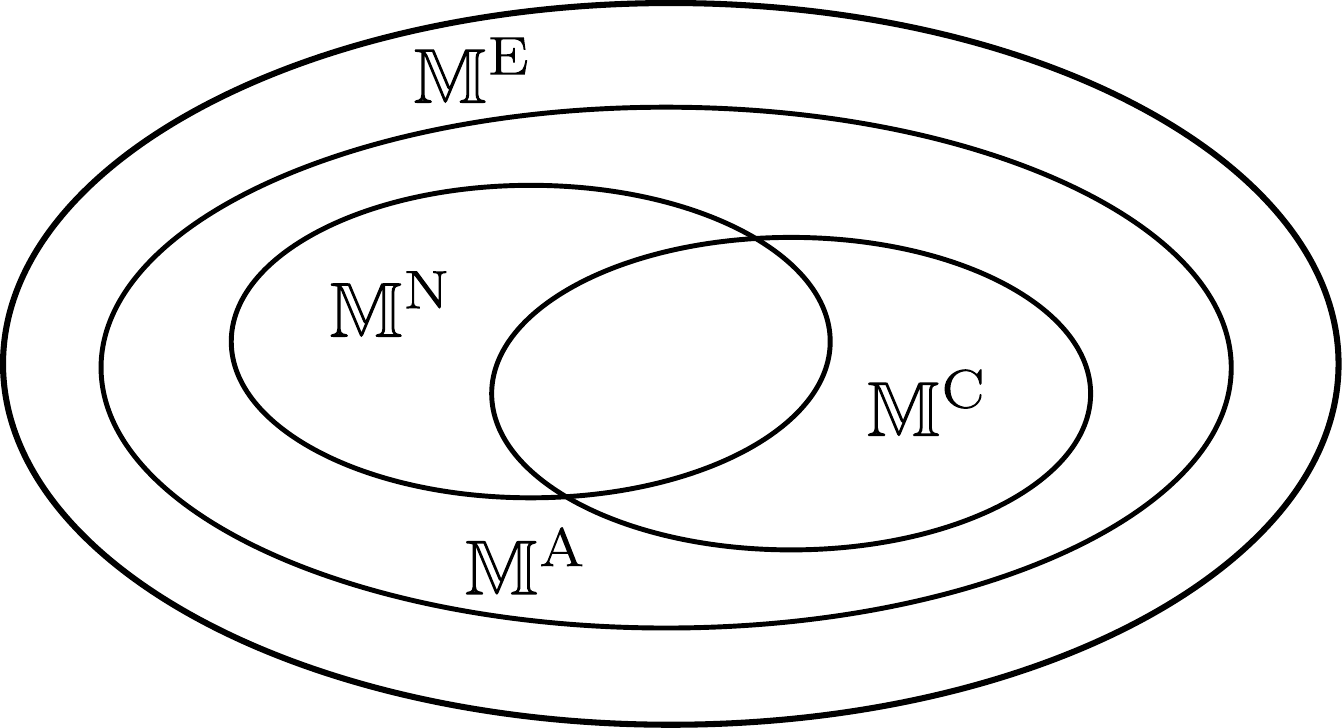}
    \caption{Relationships between classes $\metricaE$, $\metricaA$,
      $\metricaN$, and $\metricaC$.}
    \label{figura-conjuntos}
  \end{center}
\end{figure}

It follows from definitions and Fact~\ref{fato-preliminares} that
$\metricaC \subseteq \metricaA \subseteq \metricaE$, $\metricaN
\subseteq \metricaA$, $\metricaC \not\subseteq \metricaN$,
$\metricaN \not\subseteq \metricaC$, and $\metricaC \cap \metricaN \not= \emptyset$. See
Figure~\ref{figura-conjuntos}.

\section{Alignment distance}\label{sec:regular}

In this section, we describe the classes of scoring matrices that
induce $\distanciaA{\matriz}$-$p$ on sequences for each axiom $p$ of a metric. We do it through a sequence of results, such as Lemmas~\ref{distancia=0}, \ref{distanciaMaiorOuIgualAZero}, \ref{distanciaMaiorQue0}, \ref{simetriaNormal}, and   \ref{desigualdadeTriangularNormal}, which are summarized in Table~\ref{tabela2} and allow us to characterize matrices that induces each of the more general metric functions described in Section~\ref{sec:preliminares}. Moreover, as a consequence of that, we conclude this section with the important result previously stated in Section~\ref{sec:preliminares} and depicted by the following:

\begin{theorem} \label{theo:align} \rm
Let $\alphabet$ be an alphabet and $\matriz$ be a scoring matrix.
Then $\distanciaA{\matriz}$ is a metric on $\alphabet^{*}$ if and only
if 
\begin{enumerate}
\item[(\textit{i})] $\pont{\matriz}{\A}{\espaco} =
  \pont{\matriz}{\espaco}{\A} > 0$\,,
\item[(\textit{ii})] $\pont{\matriz}{\A}{\B} > 0$ if $\A \not= \B$ and
  $\pont{\matriz}{\A}{\B} = 0$ if $\A = \B$\,,
\item[(\textit{iii})] if $\pont{\matriz}{\A}{\B} <
  \pont{\matriz}{\A}{\espaco} + \pont{\matriz}{\espaco}{\B}$, then
  $\pont{\matriz}{\A}{\B} = \pont{\matriz}{\B}{\A}$\,,
\item[(\textit{iv})] $\pont{\matriz}{\A}{\espaco} \le
  \pont{\matriz}{\A}{\B} + \pont{\matriz}{\B}{\espaco}$\,, and
\item[$(v)$] $\min\{ \pont{\matriz}{\A}{\C},
  \pont{\matriz}{\A}{\espaco} + \pont{\matriz}{\espaco}{\C} \} \le
  \pont{\matriz}{\A}{\B} + \pont{\matriz}{\B}{\C}$\,,
\end{enumerate}
for each $\A, \B, \C \in \alphabet$.
\end{theorem}

We need a sequence of auxiliary results, as seen below, to eventually prove this theorem at the end of the section. 

\begin{fact}\label{fciclonegativo}\rm
Let $\matriz$ be a scoring matrix. If $D(\matriz)$ has a negative
cycle, then there exists $s \in \alphabet^{*}$ such that
$\distanciaA{\matriz}(s, s) < 0$.
\end{fact}

\begin{proof}
Suppose that $D (\matriz)$ has a cycle $C = \A_{0}, \ldots,
\A_{m}$, $\A_i \in \Sigma_{\espaco}$ such that
$\A_0 = \A_m \not= \espaco$,
$\A_{i} \not= \A_{j}$ for each
$i, j > 0$ and $\cost(C) = -X < 0$.
Let $n$ be a nonnegative integer such that $n >
(\pont{\matriz}{\A_{1}}{\espaco} +
\pont{\matriz}{\espaco}{\A_{1}})/X$ and 
$s = (\A_0 \A_1 \ldots \A_m)^n \A_0$.
Therefore,
\[
\distanciaA{\matriz}(s, s) \le \custoA{\matriz} \alinhamentoB{
  \begin{array}{ccc}
    \begin{array}{c}
      \espaco\\ \A_{0}
    \end{array} &
    \Big(
    \begin{array}{ccccc}
      \A_{0} & \A_{1} & \ldots & \A_{m - 1} & \A_{m} \\
      \A_{1} & \A_{2} & \ldots & \A_{m} & \A_{0} 
    \end{array} 
    \Big)^{n} &
    \begin{array}{c}
      \A_{0} \\ \espaco
    \end{array}\\
  \end{array}
}  <  0\,.
\]
\end{proof}

Observe that, for each $\A \in \alphabet$, we have $\custoA{\matriz}\alinhamento{\A,
  \espaco} = \pont{\matriz}{\A}{\espaco}$,
$\custoA{\matriz}\alinhamento{\espaco, \A} =
\pont{\matriz}{\espaco}{\A}$. Besides that, the only alignments of $\A, \seqVazia$
and $\seqVazia, \A$ are $\alinhamento{\A, \espaco}$ and
$\alinhamento{\espaco, \A}$, respectively. Thus, we have that
\begin{align}
  \distanciaA{\matriz}(\A, \seqVazia) &=
  \custoA{\matriz}\alinhamento{\A, \espaco} =
  \pont{\matriz}{\A}{\espaco}\,, \label{basica1}\\
  \distanciaA{\matriz}(\seqVazia, \A) &=
  \custoA{\matriz}\alinhamento{\espaco, \A} =
  \pont{\matriz}{\espaco}{\A}\,. \label{basica2}
\end{align}

Now, for each $\A, \B \in \alphabet$, since
\[
\cjtoAlinha{(\A, \B)} = \Big\{ \alinhamentoB{
  \begin{array}{c}
    \A\\
    \B
  \end{array}},
  \alinhamentoB{ 
    \begin{array}{cc}
      \A & \espaco\\
      \espaco & \B\\
    \end{array}}, 
  \alinhamentoB{
    \begin{array}{cc}
      \espaco & \A\\
      \B & \espaco\\
    \end{array}}
  \Big\}\,,
\]
$\custoA{\matriz}\alinhamento{\A,\B} = \pont{\matriz}{\A}{\B}$, and
$\custoA{\matriz}\alinhamento{\A\espaco,\espaco\B} =
\custoA{\matriz}\alinhamento{\espaco\A, \B\espaco} =
\pont{\matriz}{\A}{\espaco} + \pont{\matriz}{\espaco}{\B}$, we have
that
\begin{align}
  \distanciaA{\matriz}(\A,\B) &=  \min \bigg\{
  \begin{array}{rcl} 
    \custoA{\matriz}\alinhamento{\A, \B} & \!\!\! = \!\!\! &
    \pont{\matriz}{\A}{\B}\,, \\
    \custoA{\matriz}\alinhamento{\A\espaco, \espaco\B} & \!\!\! =
    \!\!\! & \pont{\matriz}{\A}{\espaco} +
    \pont{\matriz}{\espaco}{\B}\,.
  \end{array}
  \label{basica3}
\end{align}

\begin{lemma}\label{distancia=0}\rm
$\distanciaA{\matriz}(s, s) = 0$ for each $s \in \alphabet^{*}$
if and only if  
\begin{enumerate}
\item[(\textit{i})] $D(\matriz)$ has no negative cycle\,, and
\item[(\textit{ii})] $\pont{\matriz}{\A}{\A} = 0$ or
  $\pont{\matriz}{\A}{\espaco} + \pont{\matriz}{\espaco}{\A} = 0$, for
  each $\A \in \alphabet$.
\end{enumerate}
\end{lemma}
\begin{proof}
  Suppose that $\distanciaA{\matriz}(s, s) = 0$ for each $s \in \alphabet^{*}$.
  It follows from Fact~\ref{fciclonegativo} that (\textit{i}) is true;
  and from equation~(\ref{basica3}) that (\textit{ii}) is also true.

  Conversely, suppose that (\textit{i}) and (\textit{ii}) are true.
  Let $s = s(1) \cdots s(n) \in \alphabet^{*}$ and let $A$ be an
  A-optimal alignment of $s, s$. We define the weighted direct multigraph $H$ from
  alignment $A$ such as
  \begin{align*}
    V(H) &= \Sigma_{\,\espaco}\,, \\
    E(H) &=   \{k : [\A, \B]~\text{is aligned in $k$-th column of $A$} \}\,,\\
    \cost(\A, \B) &= \pont{\matriz}{\A}{\B}~\text{for each arc in $H$}.
\end{align*}

  By construction, $\custoA{\matriz}[A] = \cost(H)$.
  Notice that the indegree and outdegree of each vertex of $H$ are the same, which implies that 
  $H$ is an Eulerian graph
  and then $E$ can be decomposed into arc-disjoint cycles.
  For each cycle of this decomposition, there
  exists a cycle in $D(\matriz)$ with the same weight. Since, by
  hypothesis, $D(\matriz)$ has no negative cycle, it follows that
  $\cost(H) \ge 0$.

Therefore,
\begin{equation}
\distanciaA{\matriz}(s, s) = \custoA{\matriz} [A] = \cost(H) \ge 0\,.
\label{distancia=0eq1}
\end{equation}

Moreover, we are able to construct the following alignment $B$. Align
symbols $s(i), s(i)$ if $\pont{\matriz}{s(i)}{s(i)} = 0$, and align
$s(i), \espaco$ and $\espaco, s(i)$ otherwise. Since (\textit{ii})
is true, it follows that $\custoA{\matriz}[B] = 0$. Thus,
\begin{equation}
\distanciaA{\matriz}(s, s) \le \custoA{\matriz}[B] =
0\,. \label{distancia=0eq2}
\end{equation}

Using expressions~(\ref{distancia=0eq1}) and~(\ref{distancia=0eq2}),
and since the argument is used for any $s \in \alphabet^{*}$, we
conclude that $\distanciaA{\matriz}(s, s) = 0$ for each $s \in
\alphabet^{*}$.
\end{proof}

\begin{lemma}\label{distanciaMaiorOuIgualAZero}\rm
  $\distanciaA{\matriz}(s, t) \ge 0$ for each $s, t \in
  \alphabet^{*}$ if and only if for each $\A, \B \in \alphabet$,  $\pont{\matriz}{\A}{\espaco}, \pont{\matriz}{\espaco}{\B},
  \pont{\matriz}{\A}{\B} \ge 0$.
\end{lemma}
\begin{proof}
Consider $\distanciaA{\matriz}(s, t) \ge 0$ for each $s, t \in
\alphabet^{*}$. It follows from
equations~(\ref{basica1}),~(\ref{basica2}), and~(\ref{basica3}) that
$\pont{\matriz}{\A}{\espaco}  = \distanciaA{\matriz}(\A, \seqVazia) \ge 0,
\pont{\matriz}{\espaco}{\B} = \distanciaA{\matriz}(\seqVazia, \B) \ge 0$, and $\pont{\matriz}{\A}{\B} \ge \distanciaA{\matriz}(\A, \B) \ge 0$.

Conversely, consider $\pont{\matriz}{\A}{\espaco},
\pont{\matriz}{\espaco}{\B}, \pont{\matriz}{\A}{\B} \ge 0$, for each
$\A, \B \in \alphabet$. Let $A$ be an A-optimal alignment of $s,
t$. Since $\custoA{\matriz}[A]$ is the sum of entries of $\matriz$
that, by definition, are nonnegative, we have $\distanciaA{\matriz}(s,
t) = \custoA{\matriz}[A] \ge 0$.
\end{proof}

\begin{lemma}\label{distanciaMaiorQue0}\rm
  $\distanciaA{\matriz}(s, t) > 0$ for each $s \not= t \in
  \alphabet^{*}$ if and only if
  \begin{enumerate}
  \item[(\textit{i})] $\pont{\matriz}{\A}{\A} \ge 0$\,, and
  \item[(\textit{ii})] $\pont{\matriz}{\A}{\B},
    \pont{\matriz}{\A}{\espaco}, \pont{\matriz}{\espaco}{\B} > 0$\,,
  \end{enumerate}
  for each $\A \not= \B \in \alphabet$.
\end{lemma}
\begin{proof}
Consider first that $\distanciaA{\matriz}(s, t) > 0$ for each $s \not=
t \in \alphabet^{*}$. Since, for any $n \ge 0$, $\alinhamento{\A^{n +
    1}, \A^{n} \espaco}$ is an alignment of $(\A^{n + 1}, \A^{n})$
and, by hypothesis, $\distanciaA{\matriz}(\A^{n + 1}, \A^{n}) > 0$, we
have
\[
n \pont{\matriz}{\A}{\A} + \pont{\matriz}{\A}{\espaco} = 
\custoA{\matriz}  \alinhamentoB{
  \begin{array}{cc}
    \A^{n} & \A \\
    \A^{n} & \espaco
  \end{array} }
 \ge \distanciaA{\matriz}(\A^{n + 1}, \A^{n}) > 0\,.
\]
Since the expression above is valid for any $n$, this implies that
$\pont{\matriz}{\A}{\A} \ge 0$. Thus, (\textit{i}) is true. A similar
argument used in the first paragraph of proof of
Lemma~\ref{distanciaMaiorOuIgualAZero} shows that (\textit{ii}) is
also true.

Conversely, consider that (\textit{i}) and (\textit{ii}) are true.
Let $\alinhamento{s', t'}$ be an A-optimal alignment of $s, t$, $s
\not= t$. Since $s \not= t$, there exists $h$ such that $s'(h) \not=
t'(h)$, which implies by (\textit{ii}) that
$\pont{\matriz}{s'(h)}{t'(h)} > 0$. By (\textit{i}) and (\textit{ii}),
$\pont{\matriz}{s'(j)}{t'(j)} \ge 0$ for each $j \not= h$ and, by
hypothesis, $\distanciaA{\matriz}(s, t) =
\custoA{\matriz}\alinhamento{s', t'}$. Then, it follows that
\[
\distanciaA{\matriz}(s, t) = \custoA{\matriz}\alinhamento{s', t'} =
\pont{\matriz}{s'(h)}{t'(h)} + \sum_{j \not= h}
\pont{\matriz}{s'(j)}{t'(j)} 
\ge \pont{\matriz}{s'(h)}{t'(h)}
> 0\,.
\]
\end{proof}

\begin{lemma}\label{simetriaNormal}\rm
$\distanciaA{\matriz}(s, t) = \distanciaA{\matriz}(t, s)$ 
for each $s, t \in \alphabet^{*}$
if and only if
\begin{enumerate}
\item[(\textit{i})] $\pont{\matriz}{\A}{\espaco} =
  \pont{\matriz}{\espaco}{\A}$\,, and
\item[(\textit{ii})] if $\pont{\matriz}{\A}{\B} <
  \pont{\matriz}{\A}{\espaco} + \pont{\matriz}{\espaco}{\B}$, then
  $\pont{\matriz}{\A}{\B} = \pont{\matriz}{\B}{\A}$\,,
\end{enumerate}
for each $\A, \B \in \alphabet$.
\end{lemma}
\begin{proof}
Consider that $\distanciaA{\matriz}(s, t) = \distanciaA{\matriz}(t,
s)$ for each $s, t \in \alphabet^{*}$. It follows as a consequence
of equations~(\ref{basica1}) and~(\ref{basica2}) that
\[
\pont{\matriz}{\A}{\espaco} = \distanciaA{\matriz}(\A, \seqVazia) =
\distanciaA{\matriz}(\seqVazia, \A) = \pont{\matriz}{\espaco}{\A}\,.
\]
Thus, (\textit{i}) is true.

In order to check (\textit{ii}), suppose that $\pont{\matriz}{\A}{\B}
< \pont{\matriz}{\A}{\espaco} + \pont{\matriz}{\espaco}{\B}$. This
implies that
$
\custoA{\matriz}\alinhamento{\A, \B} = \pont{\matriz}{\A}{\B} < 
\pont{\matriz}{\A}{\espaco} + \pont{\matriz}{\espaco}{\B} =
\custoA{\matriz}\alinhamento{\A \espaco, \espaco \B} =
\custoA{\matriz}\alinhamento{\espaco \A, \B \espaco}\,.
$
Thus, from~(\ref{basica3}), it follows that $\distanciaA{\gamma}(\A,\B) =
\custoA{\gamma}\alinhamento{\A, \B}$. Furthermore, since
$\distanciaA{\matriz}(\B, \A) = \distanciaA{\matriz}(\A, \B)$ and,
by~$(i)$, $\pont{\matriz}{\A}{\espaco} = \pont{\matriz}{\espaco}{\A}$
and $\pont{\matriz}{\B}{\espaco} = \pont{\matriz}{\espaco}{\B}$, we
have that
\begin{align}
\distanciaA{\matriz}(\B, \A) &= \distanciaA{\matriz}(\A, \B)
= \custoA{\matriz}\alinhamento{\A,\B} = 
\pont{\matriz}{\A}{\B}\label{simetriaNormal1}
\\
&< \pont{\matriz}{\A}{\espaco} + \pont{\matriz}{\espaco}{\B} =
\pont{\matriz}{\B}{\espaco} + \pont{\matriz}{\espaco}{\A} =
\custoA{\matriz}\alinhamento{\B \espaco,
  \espaco \A} = \custoA{\matriz}\alinhamento{\espaco \B, \A
  \espaco}\,,
\nonumber \end{align}
which implies that neither $\alinhamento{\B \espaco,
  \espaco \A}$ nor $\alinhamento{\espaco \B, \A \espaco}$ are
A-optimal alignments of $\B, \A$. It follows from~(\ref{basica3}) that
\begin{equation}
\distanciaA{\matriz}(\B, \A) = \custoA{\matriz}\alinhamento{\B, \A}
= \pont{\matriz}{\B}{\A}\,.\label{simetriaNormal2}
\end{equation}
Using equations~(\ref{simetriaNormal1})
and~(\ref{simetriaNormal2}), it follows that (\textit{ii}) is also true.

Conversely, consider that (\textit{i}) and (\textit{ii}) are true and assume
that $\alinhamento{s', t'}$ is an A-optimal alignment of maximum length
of $s, t$. If $s'(h)$ or $t'(h)$ is $\espaco$, we have by~(\textit{i})
that $\pont{\matriz}{s'(h)}{t'(h)} = \pont{\matriz}{t'(h)}{s'(h)}$. If
$s'(h), t'(h) \in \alphabet$ then, by the chosen alignment $[s',t']$,
we have
\begin{align*}
\pont{\matriz}{s'(h)}{t'(h)} &+ \sum_{j \not= h}
\pont{\matriz}{s'(h)}{t'(h)} = \custoA{\matriz}[s', t'] \\
&< \custoA{\matriz} \alinhamentoB{
\begin{array}{cccc}
s'(1 \ldots j -1) & s'(j) & \espaco & s'(j + 1 \ldots \tamanho{(s',
  t')})\\
t'(1 \ldots j -1) & \espaco & t'(j) & t'(j + 1 \ldots \tamanho{(s',
  t')})
\end{array}}  \\
&= \pont{\matriz}{s'(h)}{\espaco} + \pont{\matriz}{\espaco}{t'(h)} +
\sum_{j \not= h} \pont {\matriz}{s'(h)}{t'(h)}\,.
\end{align*}
It follows that $\pont{\matriz}{s'(j)}{t'(j)} <
\pont{\matriz}{s'(j)}{\espaco} + \pont{\matriz}{\espaco}{t'(j)}$, which implies from~(\textit{ii}) that $\pont{\matriz}{s'(i)}{t'(i)} =
\pont{\matriz}{t'(i)}{s'(i)}$. Thus, regardless of the symbols $s'(i)$
and $t'(i)$, we have $\pont{\matriz}{s'(i)}{t'(i)} =
\pont{\matriz}{t'(i)}{s'(i)}$ for each $i$, which implies that
$\custoA{\matriz}\alinhamento{s', t'} =
\custoA{\matriz}\alinhamento{t', s'}$. Since $\alinhamento{s', t'}$
is an A-optimal alignment of $s, t$ and $\alinhamento{t', s'}$ is an
alignment of $t, s$, it follows that
\[
\distanciaA{\matriz}(s, t) = \custoA{\matriz}\alinhamento{s',t'} =
\custoA{\matriz}\alinhamento{t', s'} \ge \distanciaA{\matriz}(t,
s)\,.
\]
Using the same arguments we can prove that $\distanciaA{\matriz}(t, s)
\ge \distanciaA{\matriz}(s, t)$, which allows us to conclude that
$\distanciaA{\matriz}(s, t) = \distanciaA{\matriz}(t, s)$.
\end{proof}

\begin{proposition}\label{desigTriangularGeral}\rm
Let $\matriz$ be a scoring matrix, $Q$ an integer, and $s, t, u \in
\alphabet^{*}$. If
\begin{enumerate}
\item[(\textit{i})] $\pont{\matriz}{\A}{\espaco} \le
  \pont{\matriz}{\A}{\B} + \pont{\matriz}{\B}{\espaco}$\,,
\item[(\textit{ii})] $\pont{\matriz}{\espaco}{\A} \le
  \pont{\matriz}{\espaco}{\B} + \pont{\matriz}{\B}{\A}$\,,
\item[(\textit{iii})] $\min \{\pont{\matriz}{\A}{\C},
  \pont{\matriz}{\A}{\espaco} + \pont{\matriz}{\espaco}{\C}\} \le
  \pont{\matriz}{\A}{\B} + \pont{\matriz}{\B}{\C}$\,, and
\item[(\textit{iv})] $\pont{\matriz}{\B}{\espaco} +
  \pont{\matriz}{\espaco}{\B} \ge Q$\,,
\end{enumerate}
for each $\A, \B, \C \in \alphabet$, then, for each alignment $A$ of
$s, t$ and each alignment $B$ of $t, u$, there exists an alignment
$C$ of $s, u$ and an integer $k \ge 0$ such that
\[
\custoA{\matriz}[A] + \custoA{\matriz}[B] \ge \custoA{\matriz}[C] +
kQ\,, \quad \tamanho{A} \le \tamanho{C} + k\,, \quad \text{and} \quad
\tamanho{B} \le \tamanho{C} + k\,.
\]  
\end{proposition}
\begin{proof}
Suppose that (\textit{i}), (\textit{ii}), (\textit{iii}),
(\textit{iv}) are true. Let $A, B$ be alignments of $s, t$ and $t,
u$, respectively. We define
\begin{align*}
  \mathcal{C}_{1} &= \left\{h : [s(h), \espaco]~\text{is aligned in $A$}\right\}\,, \\
  \mathcal{C}_{2} &= \left\{k : [\espaco, u(k)]~\text{is aligned in $B$}\right\}\,, \\
  \mathcal{C}_{3} &= \left\{j : [\espaco, t (j)]~\text{is aligned in $A$ and $[t(j), \espaco]$
    is aligned in $B$}\right\}\,, \\
  \mathcal{C}_{4} &= \left\{(h, j) :
  ~\text{$[s(h), t(j)]$ is aligned in $A$ and 
     $[t(j), \espaco]$ is aligned in $B$}
  \right\}\,,
  \\
  \mathcal{C}_{5} &= \left\{(j, k) :~\text{$[\espaco, t (j)]$ is aligned in $A$ and $[t(j), u(k)]$
    is aligned in $B$}\right\}\,, \\
  \mathcal{C}_{6} &= \Big\{\!\! \begin{array}{l}
    (h, j, k) :~\text{$[s(h), t(j)]$ is aligned in $A$, $[t(j), u(k)]$ is aligned in $B$} \!\! \\
    \text{and}~\pont{\matriz}{s(h)}{u(k)} \le
    \pont{\matriz}{s(h)}{t(j)} + \pont{\matriz}{t(j)}{u(k)}
  \end{array} \Big\}\,, \\
  \mathcal{C}_{7} &= \Big\{\!\! \begin{array}{l}
    (h, j, k) :~\text{$[s(h), t(j)]$ is aligned in $A$, $[t(j), u(k)]$ is aligned in $B$} \!\!\\
    \text{and}~\pont{\matriz}{s(h)}{u(k)} > \pont{\matriz}{s(h)}{t(j)}
    + \pont{\matriz}{t(j)}{u(k)}
  \end{array} \Big\}\,.
\end{align*}

Thus,
\begin{align*}
\custoA{\matriz}[A] = &\sum_{h \in \mathcal{C}_{1}}
\pont{\matriz}{s(h)}{\espaco} + \sum_{j \in \mathcal{C}_{3}}
\pont{\matriz}{\espaco}{t(j)} + \sum_{(h, j) \in \mathcal{C}_{4}}
\pont{\matriz}{s(h)}{t(j)} + 
\sum_{(j, k) \in \mathcal{C}_{5}} \pont{\matriz}{\espaco}{t(j)}
\\ & + \sum_{(h, j, k) \in \mathcal{C}_{6}} \pont{\matriz}{s(h)}{t(j)} +
\sum_{(h, j, k) \in \mathcal{C}_{7}}
\pont{\matriz}{s(h)}{t(j)}
\end{align*}
and
\begin{align*}
\custoA{\matriz}[B] = &\sum_{k \in \mathcal{C}_{2}}
\pont{\matriz}{\espaco}{u(k)} + \sum_{j \in \mathcal{C}_{3}}
\pont{\matriz}{t (j)}{\espaco} + \sum_{(h, j) \in \mathcal{C}_{4}}
\pont{\matriz}{t (j)}{\espaco} + 
\sum_{(j, k) \in \mathcal{C}_{5}} \pont{\matriz}{t(j)}{u(k)}\\
& + 
\sum_{(h, j, k) \in \mathcal{C}_{6}} \pont{\matriz}{t(j)}{u(k)} + \sum_{(h, j, k) \in \mathcal{C}_{7}}
\pont{\matriz}{t(j)}{u(k)}\,,
\end{align*}
which implies that 
\begin{align*}
  \custoA{\matriz}[A] &+ \custoA{\matriz}[B] =\\
  & 
  \sum_{h \in \mathcal{C}_{1}}
  \pont{\matriz}{s(h)}{\espaco} + 
  \sum_{k \in \mathcal{C}_{2}}
  \pont{\matriz}{\espaco}{u(k)} +
  \sum_{j \in \mathcal{C}_{3}} \left(\pont{\matriz}{\espaco}{t (j)} + \pont{\matriz}{t (j)}{\espaco}\right) + \\
  & \sum_{(h, j) \in \mathcal{C}_{4}} \left(
  \pont{\matriz}{s(h)}{t (j)} + \pont{\matriz}{t (j)}{\espaco} \right) + 
  \sum_{(j, k) \in \mathcal{C}_{5}} \left( \pont{\matriz}{\espaco}{t(j)}+\pont{\matriz}{t(j)}{u(k)} \right)+ \\
  &  
  \sum_{(h, j, k) \in \mathcal{C}_{6}} \left( \pont{\matriz}{s(h)}{t(j)}+\pont{\matriz}{t(j)}{u(k)}\right) + \sum_{(h, j, k) \in \mathcal{C}_{7}}
  \left(\pont{\matriz}{s(h)}{t(j)}+ \pont{\matriz}{t(j)}{u(k)} \right)\,.
\end{align*}

Define the alignment $C$ of $s, u$ such that if $(h, j, k) \in \mathcal{C}_{6}$ then it aligns 
$[s(h), u(k)]$ and, for each remaining $s(h)$ and $u(k)$, it aligns $[s(h),
\espaco]$ and $[\espaco, u(k)]$. Thus,
\begin{align*}
\custoA{\matriz}[C] = &\sum_{h \in \mathcal{C}_{1}}
\pont{\matriz}{s(h)}{\espaco} + \sum_{k \in \mathcal{C}_{2}}
\pont{\matriz}{\espaco}{u (k)} + \sum_{(h, j) \in \mathcal{C}_{4}}
\pont{\matriz}{s(h)}{\espaco} + \sum_{(j, k) \in \mathcal{C}_{5}} \pont{\matriz}{\espaco}{u
  (k)}\\& + \sum_{(h, j, k) \in \mathcal{C}_{6}} \pont{\matriz}{s(h)}{u
  (k)} + \sum_{(h, j, k) \in \mathcal{C}_{7}} \big(
\pont{\matriz}{s(h)}{\espaco} + \pont{\matriz}{\espaco}{u(k)} \big)\,.
\end{align*}

If $j \in \mathcal{C}_{3}$, then, since (\textit{iv}) is true, we have
\begin{align}
\sum_{j \in
  \mathcal{C}_{3}} \big( \pont{\matriz}{t(j)}{\espaco} +
\pont{\matriz}{\espaco}{t(j)} \big) \ge \sum_{j \in \mathcal{C}_{3}} Q =
\tamanho{\mathcal{C}_{3}}\,{Q}\,. \label{C3}
\end{align}

From (\textit{i}),~(\textit{ii}), and the definition of
$\mathcal{C}_{6}$, we have, respectively,
\begin{align}
  \sum_{(h, j) \in \mathcal{C}_{4}} \big( \pont{\matriz}{s(h)}{t(j)} +
  \pont{\matriz}{t(j)}{\espaco} \big) &\ge \sum_{(h, j) \in \mathcal{C}_{4}}
  \pont{\matriz}{s(h)}{\espaco}\,. \label{C4} \\
  \sum_{(j, k) \in \mathcal{C}_{5}} \big(
  \pont{\matriz}{\espaco}{t(j)} + \pont{\matriz}{t(j)}{u(k)} \big)
  &\ge \sum_{(j, k) \in \mathcal{C}_{5}}
  \pont{\matriz}{\espaco}{u(k)}\,. \label{C5} \\
  \sum_{(h, j, k) \in \mathcal{C}_{6}} \big(
  \pont{\matriz}{s(h)}{t(j)} + \pont{\matriz}{t(j)}{u(k)} \big)
  &\ge \sum_{(h, j, k) \in \mathcal{C}_{6}}
  \pont{\matriz}{s(h)}{u(k)}\,. \label{C6}
\end{align}

Suppose that $(h, j, k) \in \mathcal{C}_{7}$. Then, by definition of
$\mathcal{C}_{7}$, we know that $\pont{\matriz}{s(h)}{u(k)} >
\pont{\matriz}{s(h)}{t(j)} + \pont{\matriz}{t(j)}{u(k)}$. It follows
from (\textit{iii}) that $\pont{\matriz}{s(h)}{t(j)} +
\pont{\matriz}{t(j)}{u(k)} \ge \pont{\matriz}{s(h)}{\espaco} +
\pont{\matriz}{\espaco}{u(k)}$. Thus,
\begin{align}
  \sum_{(h, j, k) \in \mathcal{C}_{7}} \big(
  \pont{\matriz}{s(h)}{t(j)} + \pont{\matriz}{t(j)}{u(k)} \big)
  \ge \sum_{(h, j, k) \in \mathcal{C}_{7}} \big(
  \pont{\matriz}{s(h)}{\espaco} + \pont{\matriz}{\espaco}{u(k)}
  \big)\,. \label{C7}
\end{align}

From equations~(\ref{C3}),~(\ref{C4}),~(\ref{C5}),~(\ref{C6}),
and~(\ref{C7}), we have
\[
\custoA{\matriz}[A] + \custoA{\matriz}[B] \ge \custoA{\matriz}[C] +
\tamanho{\mathcal{C}_{3}} \, Q\,.
\]
Hence, to finish the proof, it is enough to show that $\tamanho{A} \le
\tamanho{C} + \tamanho{\mathcal{C}_{3}}$ and $\tamanho{B} \le
\tamanho{C} + \tamanho{\mathcal{C}_{3}}$.
Since 
\begin{align*}
\tamanho{A} &= \sum_{i} \tamanho{\mathcal{C}_{i}} -
\tamanho{\mathcal{C}_{2}} \le \sum_{i} \tamanho{\mathcal{C}_{i}}\,,
\\
\tamanho{B} &= \sum_{i} \tamanho{\mathcal{C}_{i}} -
\tamanho{\mathcal{C}_{1}} \le \sum_{i} \tamanho{\mathcal{C}_{i}}\,,
\\
\tamanho{C} &= \sum_{i} \tamanho{\mathcal{C}_{i}} -
\tamanho{\mathcal{C}_{3}} + \tamanho{\mathcal{C}_{7}} \ge \sum_{i}
\tamanho{\mathcal{C}_{i}} - \tamanho{\mathcal{C}_{3}}\,,
\end{align*}
we have $\tamanho{A} \le \sum_{i} \tamanho{\mathcal{C}_{i}} \le
\tamanho{C} + \tamanho{\mathcal{C}_{3}}$ and $\tamanho{B} \le \sum_{i}
\tamanho{\mathcal{C}_{i}} \le \tamanho{C} +
\tamanho{\mathcal{C}_{3}}$.
\end{proof}

\begin{lemma}\label{desigualdadeTriangularNormal}\rm
$\distanciaA{\matriz}(s, u) \le \distanciaA{\matriz}(s, t) +
  \distanciaA{\matriz}(t, u)$ for each $s, t, u \in \alphabet^{*}$
  if and only if
\begin{enumerate}
\item[(\textit{i})] $\pont{\matriz}{\A}{\espaco} \le
  \pont{\matriz}{\A}{\B} + \pont{\matriz}{\B}{\espaco}$\,,
\item[(\textit{ii})] $\pont{\matriz}{\espaco}{\A} \le
  \pont{\matriz}{\espaco}{\B} + \pont{\matriz}{\B}{\A}$\,,
\item[(\textit{iii})] $\min \{ \pont{\matriz}{\A}{\C},
  \pont{\matriz}{\A}{\espaco} + \pont{\matriz}{\espaco}{\C} \} \le
  \pont{\matriz}{\A}{\B} + \pont{\matriz}{\B}{\C}$\,, and
\item[(\textit{iv})] $\pont{\matriz}{\B}{\espaco} +
  \pont{\matriz}{\espaco}{\B} \ge 0$\,,
\end{enumerate}
for each $\A, \B, \C \in \alphabet$.
\end{lemma}
\begin{proof}
Suppose that $\distanciaA{\matriz}(s, u) \le
\distanciaA{\matriz}(s, t) + \distanciaA{\matriz}(t, u)$ for each $s,
t, u \in \alphabet^{*}$. 
It follows from equations~(\ref{basica1}),~(\ref{basica2}),
and~(\ref{basica3}) that
\begin{align*}
\pont{\matriz}{\A}{\espaco} =  \distanciaA{\matriz}(\A, \seqVazia) &\le
\distanciaA{\matriz}(\A, \B) + \distanciaA{\matriz}(\B, \espaco) \\
&\le \custoA{\matriz}\alinhamento{\A, \B} +
\custoA{\matriz}\alinhamento{\espaco, \B} = \pont{\matriz}{\A}{\B} + \pont{\matriz}{\espaco}{\B}\,,
& \\
\pont{\matriz}{\espaco}{\A} =
\distanciaA{\matriz}(\seqVazia, \A) &\le 
\distanciaA{\matriz}(\espaco, \B) + \distanciaA{\matriz}(\B, \A) \\
&\le \custoA{\matriz}\alinhamento{\espaco, \B} +
\custoA{\matriz}\alinhamento{\B, \A} = \pont{\matriz}{\espaco}{\B} + \pont{\matriz}{\B}{\A}\,,\\
  \min \{ \pont{\matriz}{\A}{\C}, \pont{\matriz}{\A}{\espaco} +
  \pont{\matriz}{\espaco}{\C} \} &= \distanciaA{\matriz}(\A, \C) \\
  &\le \distanciaA{\matriz}(\A, \B) + \distanciaA{\matriz}(\B, \C) \le \pont{\matriz}{\A}{\B} + \pont{\matriz}{\B}{\C}\,.
\end{align*}
Therefore, (\textit{i}), (\textit{ii}), (\textit{iii}) are true.

If $\pont{\matriz}{\B}{\espaco} + \pont{\matriz}{\espaco}{\B} = 0$,
(\textit{iv}) is true. Assume then that
$\pont{\matriz}{\B}{\espaco} + \pont{\matriz}{\espaco}{\B} \not= 0$.
Let 
\[
n > \frac{\distanciaA{\matriz}(\A, \C) - (\pont{\matriz}{\A}{\espaco}
  + \pont{\matriz}{\espaco}{\C})} {\pont{\matriz}{\B}{\espaco} +
  \pont{\matriz}{\espaco}{\B}}
\] 
be a positive integer. Since $\alinhamento{\A \espaco^{n}, \espaco
  \B^{n}}$ and $\alinhamento{\B^{n} \espaco, \espaco^{n} \C}$ are
alignments of $\A, \B^{n}$ and $\B^{n}, \C$, respectively, it follows
that
\begin{align*}
\distanciaA{\matriz}(\A, \C) &\le \distanciaA{\matriz}(\A, \B^{n}) +
\distanciaA{\matriz} (\B^{n}, \C) \\
&\le \custoA{\matriz}\alinhamento{\A \espaco^{n}, \espaco \B^{n}} +
\custoA {\matriz} \alinhamento{\B^{n} \espaco, \espaco^{n} \C} \\
&= \pont{\matriz}{\A}{\espaco} + n \pont{\matriz}{\espaco}{\B} + n
\pont{\matriz}{\B}{\espaco} + \pont{\matriz}{\espaco}{\C}\,,
\end{align*}
and thus, by the choice of $n$, we have that $\pont{\matriz}{\B}{\espaco} 
+ \pont{\matriz}{\espaco}{\B} > 0$, which implies that $(\textit{iv})$ is 
also true. 

Conversely, suppose that (\textit{i}), (\textit{ii}), (\textit{iii}), and
(\textit{iv}) are true.
Let $A$ and $B$ be A-optimal alignments of $s, t$ and $t, u$,
respectively. It follows from Proposition~\ref{desigTriangularGeral}
that there exist an integer $k$ and an alignment $C$ of $s, u$ such
that
\[
\custoA{\matriz}[C] \le \custoA{\matriz}[A] + \custoA{\matriz}[B] +
0\,k = \custoA{\matriz}[A] + \custoA{\matriz}[B]\,.
\]
Consequently, since $A$ and $B$ are A-optimal alignments of $s, t$
and $t, u$, and $C$ is an alignment of $s, u$, it follows that
\[
\distanciaA{\matriz}(s, u) \le \custoA{\matriz}[C] \le
\custoA{\matriz}[A] + \custoA{\matriz}[B] = \distanciaA{\matriz}(s, t)
+ \distanciaA{\matriz}(t, u)\,.
\]
\end{proof}

\begin{table}[htpb]
\begin{minipage}{\textwidth}
\begin{center}
\begin{tabular}{clcccccc}
  & & $\prametrica$ & $\semimetrica$ & $\hemimetrica$ &
  $\pseudometrica$ & $\quasimetrica$ & $\metrica$ \\ \hline & & \\
  (a) & $D(\matriz)$ has no negative cycle & \yes & \yes & \yes & \yes
  &\yes & \yes \\ & & \\
  (b) & $\pont{\matriz}{\A}{\A} = 0$ or $\pont{\matriz}{\A}{\espaco} +
  \pont{\matriz}{\espaco}{\A} = 0$ & \yes & \yes & \yes & \yes &\yes &
  \yes \\ & & \\
  (c) & $\pont{\matriz}{\A}{\espaco}, \pont{\matriz}{\espaco}{\B},
  \pont{\matriz}{\A}{\B} \ge 0$ & \yes & \yes & \yes & \yes &\yes &
  \yes \\ & & \\
  (d) & $\pont{\matriz}{\A}{\A} \ge 0$& & \yes & & &\yes & \yes \\ & &
  \\
  (e) & $\pont{\matriz}{\A}{\espaco}, \pont{\matriz}{\espaco}{\A} > 0$
  and $\pont{\matriz}{\A}{\B} > 0$ if $\A \not= \B$ & & \yes & & &\yes
  & \yes \\& & \\
  (f) & $\pont{\matriz}{\A}{\espaco} = \pont{\matriz}{\espaco}{\A}$ &
  & \yes & & \yes & & \yes \\ & & \\
  (g) &
  \begin{tabular}{l}
    if $\pont{\matriz}{\A}{\B} < \pont{\matriz}{\A}{\espaco}
    + \pont{\matriz}{\espaco}{\B}$ then \\
    $\pont{\matriz}{\A}{\B} = \pont{\matriz}{\B}{\A}$
  \end{tabular}  
  & & \yes & & \yes & & \yes \\ & & \\
  (h) & $\pont{\matriz}{\A}{\espaco} \le \pont{\matriz}{\A}{\B} +
  \pont{\matriz}{\B}{\espaco}$ & & & \yes & \yes &\yes & \yes \\ & &
  \\
  (i) & $\pont{\matriz}{\espaco}{\A} \le \pont{\matriz}{\espaco}{\B} +
  \pont{\matriz}{\B}{\A}$ & & & \yes & \yes &\yes & \yes\\ & & \\
  (j) & 
  $\min \Big\{ \begin{array}{l}
    \pont{\matriz}{\A}{\C}, \\
    \pont{\matriz}{\A}{\espaco} + \pont{\matriz}{\espaco}{\C} 
  \end{array} \Big\} 
  \le \pont{\matriz}{\A}{\B} + \pont{\matriz}{\B}{\C}$
  &  &  & \yes & \yes &\yes & \yes\\ & & \\
  (k) & $\pont{\matriz}{\B}{\espaco} + \pont{\matriz}{\espaco}{\B} \ge
  0$ & & & \yes & \yes &\yes & \yes
\end{tabular}
\end{center}
\end{minipage}
\caption{Necessary and sufficient conditions for scoring matrix
  $\matriz$ induce $\distanciaA{\gamma}\text{-}p$ on sequences. 
  Besides \emph{metric} ($\metrica$), these properties are also used to define 
  generalized metric spaces such as \emph{premetric} ($\prametrica$),
  \emph{semimetric} ($\semimetrica$), \emph{hemimetric}
  ($\hemimetrica$), \emph{pseudometric} ($\pseudometrica$), and 
  \emph{quasimetric} ($\quasimetrica$). Results are obtained using
  definitions presented in Section~\ref{sec:preliminares}, and
  Lemmas~\ref{distancia=0}, \ref{distanciaMaiorOuIgualAZero},
  \ref{distanciaMaiorQue0}, \ref{simetriaNormal}, and
  \ref{desigualdadeTriangularNormal}, for each $\A, \B, \C \in
  \alphabet$.} \label{tabela2}
\end{table}

Table~\ref{tabela2} summarizes the results of scoring matrices $\matriz$ that induce $\distanciaA{\matriz}\text{-}p$, where $p$ is a property that allows to characterize each axiom of the metric on sequences. Finally, we can prove the preeminent result of this section.

\begin{proof} (of Theorem~\ref{theo:align})

Suppose that $\distanciaA{\matriz}$ is a metric. 
Thus, all conditions in Table~\ref{tabela2} are satisfied.
From~(e) and~(f), we have that $\pont{\matriz}{\A}{\espaco} =
\pont{\matriz}{\espaco}{\A} > 0$ and therefore (\textit{i}) is true.
Since (\textit{i}) is true, we have that
$\pont{\matriz}{\A}{\espaco} + \pont{\matriz}{\espaco}{\A} \not= 0$,
which implies by~(b) that $\pont{\matriz}{\A}{\A} = 0$;
moreover, by~(e), we have that $\pont{\matriz}{\A}{\B} > 0$ if $\A \not= \B$;
it follows that (\textit{ii}) is true.
From~(g), (h), and (j), we have that (\textit{iii}), (\textit{iv}), and
(\textit{v}) are also true.
Therefore, if $\distanciaA{\matriz}$ is a
metric, conditions (\textit{i}) to (\textit{v})
are satisfied.

Conversely, suppose that the conditions (\textit{i}) to (\textit{v})
are true and, in order to prove that these are sufficient conditions
for $\distanciaA{\matriz}$ to be a metric, we check whether all 
conditions in Table~\ref{tabela2} are satisfied.
We have, by (\textit{i}) and (\textit{ii}), that the conditions (a) to
(f), and the condition (k), are satisfied.
Since 
(\textit{iii}), (\textit{iv}) and (\textit{v}) are true,
it follows that 
(g), (h), and (j) are also true. 
Then, it is enough to show that the condition 
(i) in Table~\ref{tabela2} is true.

Suppose that $\pont{\matriz}{\B}{\A} \ge 
\pont{\matriz}{\B}{\espaco} + \pont{\matriz}{\espaco}{\A}
$. 
Since (k) is true, it follows that 
\[\pont{\matriz}{\espaco}{\A}\le 
\pont{\matriz}{\espaco}{\B} + 
\pont{\matriz}{\B}{\espaco} +
\pont{\matriz}{\espaco}{\A} 
\le 
\pont{\matriz}{\espaco}{\B} + 
\pont{\matriz}{\B}{\A}\]
and the proof is done. 
Then, assume that $\pont{\matriz}{\B}{\A} <
\pont{\matriz}{\B}{\espaco} + \pont{\matriz}{\espaco}{\A}$.
It follows from (\textit{iii}) 
that 
$\pont{\matriz}{\B}{\A} =
\pont{\matriz}{\A}{\B}$,
which implies from (\textit{i}) and (\textit{iv}) that 
\[
\pont{\matriz}{\espaco}{\A}
= \pont{\matriz}{\A}{\espaco}
\le 
\pont{\matriz}{\A}{\B} + 
\pont{\matriz}{\B}{\espaco} 
=
\pont{\matriz}{\espaco}{\B} + 
\pont{\matriz}{\B}{\A}\,.\]
Therefore, if conditions (\textit{i}) to (\textit{v})
are true, then all conditions in Table~\ref{tabela2}
are satisfied, which implies that $\distanciaA{\matriz}$ is a
metric on sequences.
\end{proof}

\section{Normalized edit distance}\label{sec:normalizado}

In this section, we describe the classes of scoring matrices that
induce $\distanciaN{\matriz}$-$p$ on sequences for each axiom $p$ of a metric
when $\distanciaN{\matriz} \in \prametrica$, as can be seen in Lemmas~\ref{lemaPrametricaN}, \ref{lema-maiorQueZero}, \ref{distanciaNSimetrica}, and \ref{distanciaNTriangular}, and summarized in Table~\ref{tabela3}. They allow us to characterize matrices that induces each of the more general metric functions described in Section~\ref{sec:preliminares}. As a consequence, we present an important result previously stated in Section~\ref{sec:preliminares} as following:

\begin{theorem} \label{theo:norm} \rm
$\distanciaN{\matriz} \in \metrica$ if and only if $\matriz \in \metricaN$.
\end{theorem}

As in the previous section, we present below a sequence of auxiliary results to finally obtain a proof of Theorem~\ref{theo:norm}. 

\begin{lemma}\label{lemaPrametricaN}\rm
Let $\matriz$ be a scoring matrix.  Then $\distanciaA{\matriz} \in
\prametrica$ if and only if $\distanciaN{\matriz} \in \prametrica$.
\end{lemma}
\begin{proof}
Let $s, t$ be sequences in $\alphabet^{*}$. In order to prove this
result, we show that $\distanciaA{\matriz}(s, s) = 0$ if and only if
$\distanciaN{\matriz}(s, s) = 0$, and that $\distanciaA{\matriz}(s, t) \ge
0$ if and only if $\distanciaN{\matriz}(s, t) \ge~0$
for each $s, t \in \Sigma^*$.

Consider first that $\distanciaA{\matriz}(s, s) = 0$. If $s =
\seqVazia$, we have $\distanciaA{\matriz}(s, s) =
\distanciaN{\matriz}(s, s) = 0$ and the proof is done. Suppose then
that $s \not= \seqVazia$. Let $A$ be an A-optimal alignment of $s,
s$. Since $\distanciaA{\matriz}(s, s) = 0$, we have that
$\custoA{\matriz}[A] = 0$ and, since $s \not= \seqVazia$, we have that
$\tamanho{A} > 0$. It follows that
\begin{equation}
\distanciaN{\matriz}(s, s) \le \custoN{\matriz}[A] =
\frac{\custoA{\matriz}[A]}{\tamanho{A}} = 0\,. \label{pra3}
\end{equation}
Let $B$ be an N-optimal alignment of $s, s$. Since $s \not= \seqVazia$,
we have $\tamanho{B} > 0$, which implies, since
$\distanciaA{\matriz}(s, s) = 0$ and $\custoA{\matriz}[B] \ge
\distanciaA{\matriz}(s, s)$, that
\[
\distanciaN{\matriz}(s, s) = \frac{\custoA{\matriz}[B]}{\tamanho{B}}
\ge \frac{\distanciaA{\matriz}(s, s)}{\tamanho{B}} = 0\,.
\]
It follows from Equation~(\ref{pra3}) that $\distanciaN{\matriz}(s, s)
= 0$ and, therefore, the proof is also done when $s \not= \seqVazia$.

Similar arguments can be used to prove that $\distanciaN{\matriz}(s,
s) = 0$ implies that $\distanciaA{\matriz}(s, s) = 0$,
and 
$\distanciaA{\matriz}(s, t) \ge 0$ if and only if
$\distanciaN{\matriz}(s, t) \ge 0$.
\end{proof} 

\begin{corolary}\label{corolaryPrametrica}\rm
Let $\matriz$ be a scoring matrix and $\A, \B \in \alphabet$. Then
$\distanciaN{\matriz} \in \prametrica$ if and only if the following
conditions are true:
\begin{enumerate}
\item[(\textit{i})] $\pont{\matriz}{\A}{\A} = 0$ or
  $\pont{\matriz}{\A}{\espaco} + \pont{\matriz}{\espaco}{\A} = 0$\,, and
\item[(\textit{ii})] $\pont{\matriz}{\A}{\espaco},
  \pont{\matriz}{\espaco}{\A}, \pont{\matriz}{\A}{\B} \ge 0$\,,
\end{enumerate}
for each $\A, \B \in \Sigma$.
\end{corolary}
\begin{proof}
Suppose that $\distanciaN{\matriz} \in \prametrica$. It follows from
Lemma~\ref{lemaPrametricaN} that $\distanciaA{\matriz} \in
\prametrica$, which implies from Table~\ref{tabela2} lines (b) and
(c), that the conditions (\textit{i}) and (\textit{ii}) are true.

Conversely, suppose that the conditions (\textit{i}) and (\textit{ii})
are true.  
It follows that (b) and (c) of Table~\ref{tabela2} are true.
Since (\textit{ii}) is true, we have that $D(\matriz)$ has no negative cycle, which implies that (a) of Table~\ref{tabela2} is also true. 
It follows that $\distanciaA{\matriz} \in \prametrica$, implying from
Lemma~\ref{lemaPrametricaN} that $\distanciaN{\matriz} \in
\prametrica$.
\end{proof}

\begin{lemma}\label{lema-maiorQueZero}\rm
Let $\distanciaN{\matriz} \in \prametrica$. Then,
$\distanciaN{\matriz}(s, t) > 0$ for any $s \not= t \in \alphabet^{*}$
if and only if $\pont{\matriz}{\A}{\espaco},
\pont{\matriz}{\espaco}{\A}, \pont{\matriz}{\A}{\B} > 0$ for any $\A
\not= \B \in \alphabet$.
\end{lemma}
\begin{proof}
Let $\distanciaN{\matriz} \in \prametrica$. Suppose that
$\distanciaN{\matriz}(s, t) > 0$ for any $s \not= t \in
\alphabet^{*}$. Then,
\begin{align*}
\pont{\matriz}{\A}{\espaco} &= \custoN{\matriz}\alinhamento{\A,
  \espaco} = \distanciaN{\matriz}[\A, \seqVazia] > 0\,,\\
\pont{\matriz}{\espaco}{\A} &= \custoN{\matriz}\alinhamento{\espaco,
  \A} = \distanciaN{\matriz}(\seqVazia, \A) > 0\,,\\
\pont{\matriz}{\A}{\B} &= \custoN{\matriz}\alinhamento{\A, \B} \ge
\distanciaN{\matriz}[\A, \B] > 0\,.
\end{align*}

Conversely, suppose that $\pont{\matriz}{\A}{\espaco},
\pont{\matriz}{\espaco}{\A}, \pont{\matriz}{\A}{\B} > 0$ for any $\A
\not= \B \in \alphabet$. Since $\distanciaN{\matriz} \in \prametrica$,
from Corollary~\ref{corolaryPrametrica} we have that
$\pont{\matriz}{\A}{\A} \ge 0$. It follows from
Lemma~\ref{distanciaMaiorQue0} that $\distanciaA{\matriz}(s, t) > 0$
for $s \not= t \in \alphabet^{*}$. Let $A$ be a N-optimal alignment
of $s, t$. It follows that
\[
\distanciaN{\matriz}(s, t) = \custoN{\matriz}[A] =
\frac{\custoA{\matriz}[A]}{\tamanho{A}} \ge
\frac{\distanciaA{\matriz}(s, t)}{\tamanho{A}} > 0\,.
\] 
\end{proof}

We denote by $\Maior$ the value of $\max_{\A \in \alphabet}
\{\pont{\matriz}{\A}{\espaco}, \pont{\matriz}{\espaco}{\A}\}$ and by
$\maior \in \alphabet$ the symbol in $\alphabet$ such that $\Maior =
\max \{\pont{\matriz}{\maior}{\espaco},
\pont{\matriz}{\espaco}{\maior}\}$.

\begin{proposition}\label{propAux2}\rm
Let $s, t \in \alphabet^{*}$. Then, $\distanciaN{\matriz}(s, t) \le
\Maior$.
\end{proposition}
\begin{proof}
Notice that $\alinhamento{s \espaco^{\tamanho{t}},
  \espaco^{\tamanho{s}} t}$ is an alignment of $s, t$. Thus,
\begin{align*}
\distanciaN{\matriz}(s, t) &\le \custoN{\matriz}  \alinhamentoB{
  \begin{array}{cc}
    s & \espaco^{\tamanho{t}}\\
    \espaco^{\tamanho{s}} & t
  \end{array}
}  \le \frac{\tamanho{s} \Maior + \tamanho{t} \Maior}{\tamanho{s}
  + \tamanho{t}} = \Maior\,.
\end{align*}
\end{proof}

\begin{proposition}\label{Q=0}\rm
Let $\distanciaN{\matriz} \in \prametrica$, $\A \in \alphabet$, and
$s, t \in \alphabet^{*}$. Then,
\begin{enumerate}
\item[(\textit{i})] If $\Maior = 0$, then $\distanciaN{\matriz}(s, t)
  = \pont{\matriz}{\A}{\espaco} = \pont{\matriz}{\espaco}{\A} = 0$\,, and
\item[(\textit{ii})] If $\Maior \not= 0$, then
  $\pont{\matriz}{\maior}{\espaco} + \pont{\matriz}{\espaco}{\maior} >
  0$ and $\pont{\matriz}{\maior}{\maior} = 0$\,.
\end{enumerate}
\end{proposition}
\begin{proof}
Suppose that $\Maior = 0$.
Since $\distanciaN{\matriz} \in \prametrica$, we have that
$\distanciaN{\matriz}(s, t) \ge 0$ and, from
Proposition~\ref{propAux2}, we have that $\distanciaN{\matriz}(s, t) \le
\Maior = 0$. It follows that $\distanciaN{\matriz}(s, t) = 0$. Since
this is true for each $s, t \in \alphabet^{*}$ and $\alinhamento{\A,
  \espaco}$, $\alinhamento{\espaco, \A}$ are the only alignments of
$\A, \seqVazia$ and $\seqVazia, \A$, respectively, we have that
\begin{align*}
  \pont{\matriz}{\A}{\espaco} &= \custoA{\matriz}\alinhamento{\A,
    \espaco} = \distanciaN{\matriz}(\A,\seqVazia) = 0 \quad
  \text{and} \\
  \pont{\matriz}{\espaco}{\A} &=
  \custoA{\matriz}\alinhamento{\espaco, \A} =
  \distanciaN{\matriz}(\seqVazia, \A) = 0\,.
\end{align*}

Suppose that $\Maior \not= 0$. Since $\distanciaN{\matriz} \in
\prametrica$, we have $\min\{ \pont{\matriz}{\maior}{\espaco},
\pont{\matriz}{\espaco}{\maior} \} \ge 0$.
It follows that 
$\Maior = \max \{ \pont{\matriz}{\maior}{\espaco},
\pont{\matriz}{\espaco}{\maior} \} \ge 
\min\{ \pont{\matriz}{\maior}{\espaco},
\pont{\matriz}{\espaco}{\maior} \} \ge 0$ and since
by hypothesis $\Maior \not= 0$, we have that 
$\max \{ \pont{\matriz}{\maior}{\espaco},
\pont{\matriz}{\espaco}{\maior} \} = \Maior > 0$.
It follows that
\[
\pont{\matriz}{\maior}{\espaco} + \pont{\matriz}{\espaco}{\maior} =
\max \{ \pont{\matriz}{\maior}{\espaco},
\pont{\matriz}{\espaco}{\maior} \} + \min \{
\pont{\matriz}{\maior}{\espaco}, \pont{\matriz}{\espaco}{\maior} \} >
0\,,
\]
which also implies, from Corollary~\ref{corolaryPrametrica}, that
$\pont{\matriz}{\maior}{\maior} = 0$.
\end{proof}

\begin{fact}\label{fato1}\rm
Let $x, z, k, w$ be real numbers. If $k \ge 0$ and $w > 0$, then
\[
\frac{kx + z}{k + w} \ge \min \left\{ x, \frac{z}{w} \right\}.
\]
\end{fact}

\begin{proposition}\label{propAux}\rm
Let $\distanciaN{\matriz} \in \prametrica$ and $\Maior \not= 0$. If
\[
\pont{\matriz}{q}{\espaco} = \pont{\matriz}{\espaco}{q} \quad
\text{or} \quad (\pont{\matriz}{\A}{\espaco} \le
\pont{\matriz}{\A}{\maior} +
\pont{\matriz}{\maior}{\espaco}~\text{and}~\pont{\matriz}{\espaco}{\B}
\le \pont{\matriz}{\espaco}{\maior} + \pont{\matriz} {\maior}{\B})\,,
\]
for each $\A, \B \in \alphabet$, then there is $n_{0}$ such that, for
each integer $n \ge n_{0}$, we have that
\[
\distanciaN{\matriz} (q^{n} \A, q^{n} \B) = \min \left\{
\begin{array}{l}
\custoN{\matriz} \alinhamento{q^{n} \A, q^{n} \B} =
\frac{\pont{\matriz}{\A}{\B}}{n + 1}\,,\\
\custoN{\matriz} \alinhamento{q^{n} \A\espaco, q^{n} \espaco\B} =
\frac{\pont{\matriz}{\A}{\espaco} + \pont{\matriz}{\espaco}{\B}}{n +
  2}
\end{array} \right\}.
\]
\end{proposition}
\begin{proof}
Let $\distanciaN{\matriz} \in \prametrica$ and $\Maior \not= 0$ such
that $\pont{\matriz}{q}{\espaco} = \pont{\matriz}{\espaco}{q}$ or
$\pont{\matriz}{\A}{\espaco} \le \pont{\matriz}{\A}{\maior} +
\pont{\matriz}{\maior}{\espaco}$ and $\pont{\matriz}{\espaco}{\B} \le
\pont{\matriz}{\espaco}{\maior} + \pont{\matriz} {\maior}{\B}$, for
each $\A, \B \in \alphabet$. Define, for each $\A, \B \in \alphabet$,
\[
n > \frac{\max \{ 
  \pont{\matriz}{\A}{\B}, 
  \pont{\matriz}{\A}{\espaco}+\pont{\matriz}{\espaco}{\B},
  \pont{\matriz}{\A}{\maior}+\pont{\matriz}{\espaco}{\B} + 
  \pont{\matriz}{\maior}{\espaco},
  \pont{\matriz}{\maior}{\B}+\pont{\matriz}{\A}{\espaco} + 
  \pont{\matriz}{\espaco}{\maior}
  \} } 
{\pont{\matriz}{\maior}{\espaco} + \pont{\matriz}{\espaco}{\maior}}\,.
\]
Since $\distanciaN{\matriz} \in \prametrica$ and $\Maior \not= 0$, we have that
$\pont{\matriz}{\maior}{\espaco} + \pont{\matriz}{\espaco}{\maior} >
0$ and $\pont{\matriz}{\maior}{\maior} = 0$
from Proposition~\ref{Q=0}, and since $\pont{\matriz}{\maior}{\maior} = 0$, we have that
\[
\custoN{\matriz} \alinhamentoB{\begin{array}{cc}
    \maior^{n} & \A\\
    \maior^{n} & \B
  \end{array}
}  = \frac{\pont{\matriz}{\A}{\B}}{n + 1}
\quad \text{and} \quad
\custoN{\matriz} \alinhamentoB{ \begin{array}{ccc}
    \maior^{n} & \A & \espaco\\
    \maior^{n} & \espaco & \B\\
  \end{array}
}  = \frac{\pont{\matriz}{\A}{\espaco} +
  \pont{\matriz}{\espaco}{\B}}{n + 2}\,.
\]
Hence, in order to prove this proposition, we have to show that 
\[
\custoN{\matriz}\alinhamento{s', t'} \ge \min \left\{
\frac{\pont{\matriz}{\A}{\B}}{n + 1},
\frac{\pont{\matriz}{\A}{\espaco} + \pont{\matriz}{\espaco}{\B}}{n +
  2} \right\},
\]
for each alignment $\alinhamento{s', t'}$ of $\maior^{n} \A,
\maior^{n} \B$.

Let $k$ be the number of symbols~$\espaco$ in $s'$. It follows that the number of symbols $\espaco$ in $t'$ is also $k$ and
$\tamanho{\alinhamento{s',t'}} = k + n + 1$. We examine four cases,
covering all possible alignments of $\maior^{n}\A ,\maior^{n}\B$.

\begin{description}
\item[Case 1:] $[\A, \B]$ is aligned in $\alinhamento{s',t'}$.

In this case, $k \ge 0$. 
Since $\pont{\matriz}{\maior}{\maior} = 0$,
we have that
\begin{align}
  \custoN{\matriz}\alinhamento{s',t'} &=
  \frac{k(\pont{\matriz}{\maior}{\espaco} +
    \pont{\matriz}{\espaco}{\maior}) + \pont{\matriz}{\A}{\B}}{k +
    n+1} \notag \\ &\ge \min \Big\{ \pont{\matriz}{\maior}{\espaco} +
  \pont{\matriz}{\espaco}{\maior}, \frac{\pont{\matriz}{\A}{\B}}{n+1}
  \Big\} = \frac{\pont{\gamma}{\A}{\B}}{n+1}\,. \label{p2}
\end{align}
Since $k \ge 0$ and $n + 1 > 0$, the inequality~(\ref{p2}) follows
from Fact~\ref{fato1} and the equality follows by the choice of $n$.

\item[Case 2:] $[\A,\espaco]$ and $[\espaco, \B]$ are aligned in
  $\alinhamento{s',t'}$.

In this case, $k \ge 1$. Since $\pont{\matriz}{\maior}{\maior} = 0$,
we have that
\begin{align*}
  \custoN{\matriz}\alinhamento{s',t'} &= \frac{(k - 1)
    (\pont{\matriz}{\maior}{\espaco} +
    \pont{\matriz}{\espaco}{\maior}) + \pont{\matriz}{\A}{\espaco} +
    \pont{\matriz}{\espaco}{\B}} {(k - 1) + n + 2}\\
  &\ge \min \Big\{ \pont{\matriz}{\maior}{\espaco} +
  \pont{\matriz}{\espaco}{\maior}, \frac{\pont{\matriz}{\A}{\espaco} +
    \pont{\matriz}{\espaco}{\B}} {n + 2} \Big\} =
\frac{\pont{\matriz}{\A}{\espaco} +
  \pont{\matriz}{\espaco}{\B}}{n+2}\,.
\end{align*}
The inequality, since $k - 1 \ge 0$ and $n + 2 > 0$, follows from
Fact~\ref{fato1} and the last equality follows by the choice of $n$.

\item[Case 3:] $[\A,\maior]$ is aligned in $\alinhamento{s',t'}$. 

In this case, $k \ge 1$. Also, since $\pont{\matriz}{\maior}{\maior} =
0$, we have that
\begin{align}
  \custoN{\matriz}\alinhamento{s',t'} &= \frac{(k - 1)
    (\pont{\matriz}{\maior}{\espaco} +
    \pont{\matriz}{\espaco}{\maior}) +
    \pont{\matriz}{\A}{\maior}+\pont{\matriz}{\espaco}{\B} +
    \pont{\matriz}{\maior}{\espaco}}{(k - 1)+ n+2} \nonumber \\
  &\ge \min \Big\{ \pont{\matriz}{\maior}{\espaco} +
  \pont{\matriz}{\espaco}{\maior},
  \frac{\pont{\matriz}{\A}{\maior}+\pont{\matriz}{\espaco}{\B}+
    \pont{\matriz}{\maior}{\espaco}}{n+2} \Big\} \label{p3}\\
  &= \frac{\pont{\matriz}{\A}{\maior}+\pont{\matriz}{\espaco}{\B}+
    \pont{\matriz}{\maior}{\espaco}}{n+2}\,. \label{p4}
\end{align}
Since $k - 1 \ge 0$ and $n + 2 > 0$, inequality~(\ref{p3}) follows
from Fact~\ref{fato1} and equality~(\ref{p4}) follows 
by the choice of $n$.

Suppose that $\pont{\matriz}{\maior}{\espaco} =
\pont{\matriz}{\espaco}{\maior}$. Then,
$\pont{\matriz}{\maior}{\espaco} = \min
\{\pont{\matriz}{\maior}{\espaco},\pont{\matriz}{\espaco}{\maior} \} =
\Maior$. Since $\distanciaN{\matriz} \in \prametrica$, we have from
Corollary~\ref{corolaryPrametrica} that $\pont{\matriz}{\A}{\maior}
\ge 0$. It follows from equality~(\ref{p4}) and the definition of
$\Maior$ that
\begin{align*}
  \custoN{\matriz}\alinhamento{s',t'} &\ge
  \frac{\pont{\matriz}{\A}{\maior}+\pont{\matriz}{\espaco}{\B}+
    \pont{\matriz}{\maior}{\espaco}}{n+2} \ge 
    \frac{0 + \pont{\matriz}{\espaco}{\B}+ \Maior}{n+2} \\
    &\ge \frac{\pont{\matriz}{\espaco}{\B} +
    \pont{\matriz}{\A}{\espaco}}{n+2}
    = \frac{\pont{\matriz}{\A}{\espaco} + \pont{\matriz}{\espaco}{\B}
    }{n+2}\,,
\end{align*}
and the proof is complete. Suppose then that $\pont{\matriz}{\maior}{\espaco} \not=
\pont{\matriz}{\espaco}{\maior}$. By hypothesis, this implies that
$\pont{\matriz}{\A}{\espaco} \le \pont{\matriz}{\A}{q} +
\pont{\matriz}{q}{\espaco}$. It follows from equality~(\ref{p4}) that
\[
\custoN{\matriz}\alinhamento{s',t'} \ge
\frac{\pont{\matriz}{\A}{\maior}+\pont{\matriz}{\espaco}{\B}+
  \pont{\matriz}{\maior}{\espaco}}{n+2} \ge
\frac{\pont{\matriz}{\A}{\espaco}+\pont{\matriz}{\espaco}{\B}}{n+2}\,.
\]

\item[Case 4:] $[\maior, \B]$ is aligned in $\alinhamento{s',t'}$. Similar to Case 3.
\end{description}
\end{proof}

\begin{proposition}\label{propAux-3}\rm
Suppose that $\distanciaN{\matriz} \in \prametrica$. Let $\A, \B \in
\alphabet$ and $s, t \in \alphabet^{*}$. If $[\A, \B]$ is aligned in
an N-optimal alignment of $s, t$ of maximum length, then
\[
\pont{\matriz}{\A}{\B} < \pont{\matriz}{\A}{\espaco} +
\pont{\matriz}{\espaco}{\B}\,.
\] 
\end{proposition}
\begin{proof}
Let $\alinhamento{s', t'}$ be a N-optimal alignment of maximum length
of $s, t$ and $j$ be an integer such that 
$s'(j) = \A$ and $t'(j) = \B$. Consider the following alignment of $s, t$
\[
A =  
\alinhamentoB{ \begin{array}{cccccccc}
s'(1) & \ldots & s'(j - 1) & \A & \espaco & s'(j+1) & \ldots & 
s'(\tamanho{s'})\\
t'(1) & \ldots & t'(j - 1) & \espaco & \B & t'(j+1) & \ldots & 
t'(\tamanho{t'})
\end{array}
}.
\]

Since $\alinhamento{s',t'}$ is an N-optimal alignment
of maximum length and $\abs{[s',t']} < \abs{A}$, we have that
\begin{align*}
\frac{\custoA{\matriz}\alinhamento{s',t'}}{\tamanho{\alinhamento{s',t'}}}
= \custoN{\matriz}\alinhamento{s', t'} &< \custoN{\matriz}[A] =
\frac{\custoA{\matriz}[A]}{\tamanho{A}}\\
&= \frac{\custoA{\matriz}\alinhamento{s',t'} +
  \pont{\matriz}{\A}{\espaco} + \pont{\matriz}{\espaco}{\B} -
  \pont{\matriz}{\A}{\B}}{\tamanho{\alinhamento{s',t'}}+1}\,,
\end{align*}
which implies that 
\[
\frac{\custoA{\matriz}\alinhamento{s',t'}}{\tamanho{\alinhamento{s',t'}}}
< \pont{\matriz}{\A}{\espaco} + \pont{\matriz}{\espaco}{\B} -
\pont{\matriz}{\A}{\B}\,. 
\]
Since $\distanciaN{\matriz} \in \prametrica$, we have that
$\distanciaN{\matriz}(s, t) \ge 0$.
It follows that
\[
0 \le \distanciaN{\matriz}(s, t) =
\frac{\custoA{\matriz}\alinhamento{s',t'}}{\tamanho{\alinhamento{s',t'}}}
< \pont{\matriz}{\A}{\espaco} + \pont{\matriz}{\espaco}{\B} -
\pont{\matriz}{\A}{\B}\,,
\]
which implies that $\pont{\matriz}{\A}{\B} <
\pont{\matriz}{\A}{\espaco} + \pont{\matriz}{\espaco}{\B}$.
\end{proof}

\begin{fact}\label{fato2}\rm
Let $x, y \in \mathbb{R}$.  If $x < y$ then there exists an
integer $n_{0}$ such that
\[
\frac{x}{n + 1} < \frac{y}{n + 2}\,,
\]
for each $n > n_{0}$.
\end{fact}
 
\begin{lemma}\label{distanciaNSimetrica}\rm
Let $\distanciaN{\matriz} \in \prametrica$. Then,
$\distanciaN{\matriz}(s, t) = \distanciaN{\matriz}(t, s)$ for each $s,
t \in \alphabet^{*}$ if and only if
\begin{enumerate}
\item [(\textit{i})] $\pont{\matriz}{\A}{\espaco} =
  \pont{\matriz}{\espaco}{\A}$\,, and
\item [(\textit{ii})] if $\pont{\matriz}{\A}{\B} <
  \pont{\matriz}{\A}{\espaco} + \pont{\matriz}{\espaco}{\B}$, then
  $\pont{\matriz}{\A}{\B} = \pont{\matriz}{\B}{\A}$\,,
\end{enumerate}
for each $\A, \B \in \alphabet$.
\end{lemma}
\begin{proof}
By Corollary~\ref{corolaryPrametrica}, we have that
$\pont{\matriz}{\A}{\B}, \pont{\matriz}{\maior}{\espaco},
\pont{\matriz}{\espaco}{\maior} \ge 0$.

Suppose that $\distanciaN{\matriz}(s, t) =
\distanciaN{\matriz}(t, s)$ for each $s, t \in \alphabet^{*}$.
Since $\alinhamento{\A, \espaco}$ and $\alinhamento{\espaco, \A}$ are
the only alignments of $\A, \seqVazia$ and~$\seqVazia, \A$,
respectively, and by hypothesis, $\distanciaN{\matriz}(\A, \seqVazia) =
\distanciaN{\matriz}(\seqVazia, \A)$, we have that
\[
\pont{\matriz}{\A}{\espaco} =
\custoN{\matriz}\alinhamento{\A, \espaco} = 
\distanciaN{\matriz}(\A, \seqVazia) =  
\distanciaN{\matriz}(\seqVazia, \A) = 
\custoN{\matriz}\alinhamento{\espaco, \A} =  
\pont{\matriz}{\espaco}{\A}\,,
\]
which implies that (\textit{i}) is true.
In order to show (\textit{ii}), we consider two possibilities: $\Maior
= 0$ and $\Maior \not= 0$. If $\Maior = 0$, since
$\pont{\matriz}{\A}{\B} \ge 0$ and $\Maior \ge
\pont{\matriz}{\A}{\espaco}, \pont{\matriz}{\espaco}{\B}$, we have
that
$
\pont{\matriz}{\A}{\B} \ge 0 = 0 + 0 = \Maior + \Maior \ge 
\pont{\matriz}{\A}{\espaco} + \pont{\matriz}{\espaco}{\B}
$ and in this case 
(\textit{ii}) is vacuously satisfied. Thus, assume that $\Maior \not= 0$ and suppose that $\pont{\matriz}{\A}{\B} <
\pont{\matriz}{\A}{\espaco} + \pont{\matriz}{\espaco}{\B}$. Choose $n$
big enough satisfying both Proposition~\ref{propAux}, since $\pont{\matriz}{\maior}{\espaco} = \pont{\matriz}{\espaco}{\maior}$, and 
Fact~\ref{fato2}. It follows that
\begin{align*}
\distanciaN{\matriz}(\maior^{n} \A, \maior^{n} \B) &= 
\min \left\{ \frac{\pont{\matriz}{\A}{\B}}{n + 1}, 
\frac{\pont{\matriz}{\A}{\espaco} + \pont{\matriz}{\espaco}{\B}}{n + 2}
\right\}= \frac{\pont{\matriz}{\A}{\B}}{n + 1}
< \frac{\pont{\matriz}{\A}{\espaco} +  \pont{\matriz}{\espaco}{\B}}{n +
 2} \hspace{0.2cm} \mbox{and}\\
\distanciaN{\matriz}(\maior^{n} \B, \maior^{n} \A) &=  
\min \left\{ \frac{\pont{\matriz}{\B}{\A}}{n + 1}, 
\frac{\pont{\matriz}{\B}{\espaco} + \pont{\matriz}{\espaco}{\A}}{n + 2}
\right\}.
\end{align*}
Since $\pont{\matriz}{\A}{\espaco} =
\pont{\matriz}{\espaco}{\A}$, $\pont{\matriz}{\B}{\espaco} =
\pont{\matriz}{\espaco}{\B}$, and $\distanciaN{\matriz}(\maior^{n} \A,
\maior^{n} \B) = \distanciaN{\matriz}(\maior^{n} \B, \maior^{n}
\A)$, we have that $(\pont{\matriz}{\B}{\espaco} +
\pont{\matriz}{\espaco}{\A})/(n + 2)$ is not a value of
$\distanciaN{\matriz}(\maior^{n} \B, \maior^{n} \A)$, which implies
that $\distanciaN{\matriz}(\maior^{n} \B, \maior^{n} \A) =
\pont{\matriz}{\B}{\A}/(n + 1)$. Therefore,
\[
\frac{\pont{\matriz}{\A}{\B}}{n + 1} =
\distanciaN{\matriz}(\maior^{n} \A, \maior^{n} \B) = 
\distanciaN{\matriz}(\maior^{n} \B, \maior^{n} \A) = 
\frac{\pont{\matriz}{\B}{\A}}{n + 1}\,,
\]
and thus $\pont{\matriz}{\A}{\B} = \pont{\matriz}{\B}{\A}$.

Conversely, consider that (\textit{i}) and (\textit{ii}) are true.
Let $\alinhamento{s', t'}$ be an N-optimal alignment of maximum length
of $s, t$. If $s'(j) = \espaco$ or $t'(j) = \espaco$ then from
(\textit{i}) we have that $\pont{\matriz}{s'(j)}{t'(j)} =
\pont{\matriz}{t'(j)}{s'(j)}$. If $s'(j) \not= \espaco$ and $t'(j)
\not= \espaco$ then, since 
$\distanciaN{\matriz} \in \prametrica$, we have from
Proposition~\ref{propAux-3} that $\pont{\matriz}{s'(j)}{t'(j)} <
\pont{\matriz}{s'(j)}{\espaco} + \pont{\matriz}{\espaco}{t'(j)}$,
which implies by (\textit{ii}) that $\pont{\matriz}{s'(j)}{t'(j)} =
\pont{\matriz}{t'(j)}{s'(j)}$. Using these observations, we have that
\begin{align*}
\distanciaN{\matriz}(s, t) &= \custoN{\matriz}\alinhamento{s', t'} =
\frac{\sum_{j} \pont{\matriz}{s'(j)}{t'(j)}}{\tamanho{(s', t')}} \\
&= \frac{\sum_{j} \pont{\matriz}{t'(j)}{s'(j)}}{\tamanho{(t', s')}} =
\custoN{\matriz}\alinhamento{t', s'} \ge \distanciaN{\matriz}(t,
s)\,.
\end{align*}
Similarly, we have $\distanciaN{\matriz}(s, t) \le
\distanciaN{\matriz}(t, s)$, which allows us to conclude that
$\distanciaN{\matriz}(s, t) = \distanciaN{\matriz}(t, s)$.
\end{proof}

\begin{proposition}\label{AuxdistanciaNTriangular-1}\rm
Let $\distanciaN{\matriz} \in \prametrica$.  If $\distanciaN{\matriz}(s, t) \le \distanciaN{\matriz}
(s, u) + \distanciaN{\matriz} (u, t)$ for all $s, t, u \in
\alphabet^{*}$ then
\[
\pont{\matriz}{\A}{\espaco} \le 
\pont{\matriz}{\A}{\B} + \pont{\matriz}{\B}{\espaco}
\qquad \text{and} \qquad
\pont{\matriz}{\espaco}{\A} \le
\pont{\matriz}{\espaco}{\B} + \pont{\matriz}{\B}{\A}\,,
\]
for all $\A, \B \in
\alphabet$.
\end{proposition}
\begin{proof}
Suppose that $\distanciaN{\matriz}(s, t) \le \distanciaN{\matriz}(s, u) +
\distanciaN{\matriz}(u, t)$ for all $s, t, u \in \alphabet^{*}$.
Since $\alinhamento{\A, \espaco}$ is the only alignment of $\A,
\seqVazia$, we have that
$
\distanciaN{\matriz}(\A, \seqVazia) = 
\pont{\matriz}{\A}{\espaco}$, and since $\alinhamento{\A, \B}$ and $\alinhamento{\B, \espaco}$ are
alignments of $\A, \B$ and $\B, \seqVazia$, it follows
that
\begin{align*}
\pont{\matriz}{\A}{\espaco} = \distanciaN{\matriz}(\A, \seqVazia) &\le
\distanciaN{\matriz}(\A, \B) + \distanciaN{\matriz}(\B, \seqVazia) \\
&\le \custoN{\matriz}\alinhamento{\A, \B} +
\custoN{\matriz}\alinhamento{\B, \espaco} \\
&= \pont{\matriz}{\A}{\B} + \pont{\matriz}{\B}{\espaco}\,.
\end{align*}

Using similar arguments we also prove that
$\pont{\matriz}{\espaco}{\A} \le \pont{\matriz}{\espaco}{\B} +
\pont{\matriz}{\B}{\A}$.
\end{proof}

\begin{proposition}\label{AuxdistanciaNTriangular}\rm
Let $\distanciaN{\matriz} \in \prametrica$. If
$\distanciaN{\matriz}(s, t) \le \distanciaN{\matriz}(s, u) +
\distanciaN{\matriz}(u, t)$ for each $s, t, u \in \alphabet^{*}$
then
\[
\max\{\pont{\matriz}{\A}{\espaco}, \pont{\matriz}{\espaco}{\A}\}
\le \pont{\matriz}{\B}{\espaco} + \pont{\matriz}{\espaco}{\B}\,,
\]
for each and $\A, \B \in \alphabet$.
\end{proposition}
\begin{proof}
Suppose that
$\distanciaN{\matriz}(s, t) \le \distanciaN{\matriz}(s, u) +
\distanciaN{\matriz}(u, t)$ for all $s, t, u \in \alphabet^{*}$.

Since $\distanciaN{\matriz} \in \prametrica$,
if $\Maior = 0$, then
$\pont{\matriz}{\A}{\espaco} = \pont{\matriz}{\espaco}{\A} =
\pont{\matriz}{\B}{\espaco} = 
\pont{\matriz}{\espaco}{\B} = 
0$ as a consequence of Corollary~\ref{corolaryPrametrica}
and, therefore, the propositions is proved. 
Thus, assume that $\Maior \not= 0$. That is,
$\Maior > 0$ from Corollary~\ref{corolaryPrametrica}.
W.l.o.g., assume that 
$\pont{\matriz}{\maior}{\espaco} = \Maior$.

Suppose that there exists $\A, \B \in \alphabet$
such that $\max\{\pont{\matriz}{\A}{\espaco}, \pont{\matriz}{\espaco}{\A}\}
> \pont{\matriz}{\B}{\espaco} +
\pont{\matriz}{\espaco}{\B}$ by contradiction. This implies that
$\Maior > \pont{\matriz}{\B}{\espaco} +
\pont{\matriz}{\espaco}{\B}$. Let $k$ be an positive integer such that
\[
k > \frac{\pont{\matriz}{\B}{\espaco}+\pont{\matriz}{\espaco}{\B}}
{\Maior - (\pont{\matriz}{\B}{\espaco} +
  \pont{\matriz}{\espaco}{\B})}\,.
\]

Since $\distanciaN{\matriz}
\in \prametrica$, we have that
$\pont{\matriz}{\maior}{\espaco}, \pont{\matriz}{\espaco}{\maior} \ge 0$,
which implies, since 
$\pont{\matriz}{\maior}{\espaco} = \Maior >0$, that $\pont{\matriz}{\maior}{\espaco} + \pont{\matriz}{\espaco}{\maior} > 0$,
which in turn implies that  $\pont{\matriz}{\maior}{\maior} = 0$
as a consequence of
Corollary~\ref{corolaryPrametrica}. Then,
\begin{align}
\distanciaN{\matriz}(\maior^{k} ,\maior^{k} \B) &\le \custoN{\matriz} 
\alinhamentoB{
\begin{array}{cc}
  \maior^{k} & \espaco \\
  \maior^{k} & \B
\end{array}
} =
\frac{\pont{\matriz}{\espaco}{\B}}{k+1}\,, \label{AuxdistanciaNTriangular2}\\
\distanciaN{\matriz}(\maior^{k} \B, \B) &\le \custoN{\matriz} 
\alinhamentoB{
\begin{array}{ccc}
  \maior^{k} & \B & \espaco \\
\espaco^{k} & \espaco & \B
\end{array}
} = 
\frac{k \pont{\matriz}{\maior}{\espaco} + (\pont{\matriz}{\B}{\espaco} +
  \pont{\matriz}{\espaco}{\B})}{k+2}\,.\label{AuxdistanciaNTriangular3}
\end{align} 

In any alignment of $\maior^{k}, \B$ either $[\maior, \B]$ or $[\espaco, \B]$ is aligned. Consequently,
\begin{align*}
\distanciaN{\matriz}(\maior^{k}, \B) &= 
\min \left\{ \custoN{\matriz} 
\alinhamentoB{
  \begin{array}{cc}
    \maior^{k-1} & \maior \\
    \espaco^{k-1} & \B
  \end{array}
},
\custoN{\matriz}
\alinhamentoB{
  \begin{array}{cc}
    \maior^{k} & \espaco \\
    \espaco^{k} & \B
  \end{array}
}
\right\} \\
&= 
\min \left\{ \frac{(k-1) \pont{\matriz}{\maior}{\espaco} +
  \pont{\matriz}{\maior}{\B}}{k}, \frac{k \pont{\matriz}{\maior}{\espaco} +
  \pont{\matriz}{\espaco}{\B}}{k+1} \right\}\,.
\end{align*}
Suppose that $\distanciaN{\gamma}(\maior^k,\B) = 
{((k-1) \pont{\matriz}{\maior}{\espaco} +
  \pont{\matriz}{\maior}{\B})}/{k}$. 
It follows that 
\[
\frac{(k-1) \pont{\matriz}{\maior}{\espaco} + \pont{\matriz}{\maior}{\B}}{k}
\le \frac{k \pont{\matriz}{\maior}{\espaco} +
  \pont{\matriz}{\espaco}{\B}}{k+1}\,.
\]
We
have that $\pont{\matriz}{\maior}{\espaco} \le \pont{\matriz}{\maior}{\B} +
\pont{\matriz}{\B}{\espaco}$ by Proposition~\ref{AuxdistanciaNTriangular-1}, which implies, since $k > 0$,  that
\[
\frac{k \pont{\matriz}{\maior}{\espaco} - \pont{\matriz}{\B}{\espaco}}{k}
\le \frac{(k-1) \pont{\matriz}{\maior}{\espaco} +
  \pont{\matriz}{\maior}{\B}}{k}\,.
\]
It follows that
\[
\frac{k \pont{\matriz}{\maior}{\espaco} - \pont{\matriz}{\B}{\espaco}}{k}
\le \frac{k \pont{\matriz}{\maior}{\espaco} +
  \pont{\matriz}{\espaco}{\B}}{k+1}\,,
\]
and, since $k>0$, that
$k(\pont{\gamma}{\maior}{\espaco} -(\pont{\matriz}{\B}{\espaco} +
\pont{\matriz}{\espaco}{\B})) \le  \pont{\matriz}{\B}{\espaco}$
which implies, since $\Maior =
\pont{\matriz}{\maior}{\espaco} > \pont{\matriz}{\B}{\espaco} +
\pont{\matriz}{\espaco}{\B}$, that $k \le \pont{\matriz}{\B}{\espaco}/
(\Maior - (\pont{\matriz}{\B}{\espaco} +
\pont{\matriz}{\espaco}{\B}))$, contradicting the choice of
$k$ since $\pont{\matriz}{\espaco}{\B} \ge 0$. Thus,
we assume that
\begin{equation}
\distanciaN{\matriz}(\maior^{k}, \B) = \frac{k \pont{\matriz}{\maior}{\espaco}
  + \pont{\matriz}{\espaco}{\B}}{k+1}\,.
\label{AuxdistanciaNTriangular4}
\end{equation}

Since $\distanciaN{\matriz}(\maior^{k}, \B) \le
\distanciaN{\matriz}(\maior^{k} ,\maior^{k} \B) + \distanciaN{\matriz}(\maior^{k}
\B, \B)$, we have from Equations~(\ref{AuxdistanciaNTriangular2}),
(\ref{AuxdistanciaNTriangular3}), and~(\ref{AuxdistanciaNTriangular4})
that
\begin{align*}
\frac{k \pont{\matriz}{\maior}{\espaco} +
  \pont{\matriz}{\espaco}{\B}}{k+1} = \distanciaN{\matriz}(\maior^{k}, \B)
&\le \distanciaN{\matriz}(\maior^{k} ,\maior^{k} \B) +
\distanciaN{\matriz}(\maior^{k} \B, \B) \\
&\le \frac{\pont{\matriz}{\espaco}{\B}}{k+1} + \frac{k
  \pont{\matriz}{\maior}{\espaco} + (\pont{\matriz}{\B}{\espaco} +
  \pont{\matriz}{\espaco}{\B})}{k+2}\,,
\end{align*}
which implies, since $k > 0$ and $\Maior = \pont{\matriz}{\maior}{\espaco} >
\pont{\matriz}{\B}{\espaco} + \pont{\matriz}{\espaco}{\B}$, that
\[
k \le \frac{\pont{\matriz}{\B}{\espaco} + \pont{\matriz}{\espaco}{\B}}
{\Maior - (\pont{\matriz}{\B}{\espaco} +
  \pont{\matriz}{\espaco}{\B})}\,,
\]
contradicting again the choice of $k$. Thus, 
$\max\{\pont{\matriz}{\A}{\espaco}, \pont{\matriz}{\espaco}{\A}\}
\le \pont{\matriz}{\B}{\espaco} + \pont{\matriz}{\espaco}{\B}$
for each $\A,\B~\in~\Sigma$. 
\end{proof}

\begin{proposition}\label{AuxdistanciaNTriangular-2}\rm
Let $\distanciaN{\matriz} \in \prametrica$. If $\distanciaN{\matriz}(s, t) \le
\distanciaN{\matriz}(s, u) + \distanciaN{\matriz}(u, t)$ for each $s,
t, u \in \alphabet^{*}$ then
\[
\min \{ \pont{\matriz}{\A}{\C}, \pont{\matriz}{\A}{\espaco} +
\pont{\matriz}{\espaco}{\C} \} \le \pont{\matriz}{\A}{\B} +
\pont{\matriz}{\B}{\C}\,,
\]
for each $\A, \B, \C \in \alphabet$.
\end{proposition}
\begin{proof}
Since $\distanciaN{\gamma} \in \prametrica$, from  Corollary~\ref{corolaryPrametrica} we have that
$\pont{\gamma}{\A}{\B}, \pont{\gamma}{\A}{\C}, \pont{\gamma}{\B}{\C} \ge 0$ for each $\A, \B, \C \in \Sigma$.

If $\mathcal{Q} = 0$, from Proposition~\ref{Q=0}, we have 
$\pont{\gamma}{\A}{\espaco} = \pont{\gamma}{\espaco}{\A} = 0$
for each $\A \in \Sigma$ since $\distanciaN{\gamma} \in \prametrica$.
Since $\pont{\gamma}{\A}{\B}, \pont{\gamma}{\A}{\C}, \pont{\gamma}{\B}{\C} \ge 0$ for each $\A, \B, \C \in \Sigma$,
it follows that 
$\min \{ \pont{\matriz}{\A}{\C}, \pont{\matriz}{\A}{\espaco} +
\pont{\matriz}{\espaco}{\C} \} = 0 \le \pont{\matriz}{\A}{\B} +
\pont{\matriz}{\B}{\C}$ and the proposition is proved. 
Thus, assume that $\mathcal{Q} \not= 0$.

Suppose that
$\distanciaN{\matriz}(s, t) \le \distanciaN{\matriz}(s, u) +
\distanciaN{\matriz}(u, t)$ for each $s, t, u \in\alphabet^{*}$. 
We have that
$\pont{\matriz}{\A}{\espaco} \le \pont{\matriz}{\A}{\maior} +
\pont{\matriz}{\maior}{\espaco}$ 
and 
$\pont{\matriz}{\espaco}{\C} \le \pont{\matriz}{\espaco}{\maior} +
\pont{\matriz}{\maior}{\C}$ 
for each $\A, \C \in \alphabet$,
from
Proposition~\ref{AuxdistanciaNTriangular-1}. Let
$n_{0}$ be an integer satisfying Proposition~\ref{propAux}
and Fact~\ref{fato2}. Then, for $n \ge n_{0}$, it follows that
\begin{align*}
\min \left\{ 
\frac{\pont{\matriz}{\A}{\C}}{n+1},
\frac{\pont{\matriz}{\A}{\espaco} + \pont{\matriz}{\espaco}{\C}}{n+2}
\right\} &= \distanciaN{\matriz}(\maior^{n}\A, \maior^{n}\C) \\
 &\le \distanciaN{\matriz}(\maior^{n}\A, \maior^{n}\B) 
+ \distanciaN{\matriz}(\maior^{n}\B, \maior^{n}\C) \\
&\le \frac{\pont{\matriz}{\A}{\B}}{n+1} +
\frac{\pont{\matriz}{\B}{\C}}{n+1}\,.
\end{align*}
This implies that 
if $\distanciaN{\gamma}(q^n \A,q^n \C) = 
\pont{\gamma}{\A}{\C}/(n+1)$, then 
$\pont{\gamma}{\A}{\C} \le \pont{\gamma}{\A}{\B} + \pont{\gamma}{\B}{\C}$ and the proposition is proved. 
Then, assume that
\[
\frac{\pont{\matriz}{\A}{\espaco} + \pont{\matriz}{\espaco}{\C}}{n+2}
= \min \left\{ \frac{\pont{\matriz}{\A}{\C}}{n+1},
\frac{\pont{\matriz}{\A}{\espaco} + \pont{\matriz}{\espaco}{\C}}{n+2}
\right\} \le \frac{\pont{\matriz}{\A}{\B}}{n+1} +
\frac{\pont{\matriz}{\B}{\C}}{n+1}\,.
\]
It follows from contrapositive of Fact~\ref{fato2} that 
$\pont{\gamma}{\A}{\espaco} + \pont{\gamma}{\espaco}{\A} \le \pont{\gamma}{\A}{\B}+\pont{\gamma}{\B}{\C}$ since $n$ is big enough and the proposition is proved.

\end{proof}

\begin{lemma}\label{distanciaNTriangular}\rm
Let $\distanciaN{\matriz} \in \prametrica$. Then,
$\distanciaN{\matriz}(s, t) \le \distanciaN{\matriz} (s, u) +
\distanciaN{\matriz} (u, t)$ for each $s, t, u \in \alphabet^{*}$ if
and only if
\begin{enumerate}
\item [(\textit{i})] $\pont{\matriz}{\A}{\espaco} \le
  \pont{\matriz}{\A}{\B} + \pont{\matriz}{\B}{\espaco}$\,,
\item [(\textit{ii})] $\pont{\matriz}{\espaco}{\A} \le
  \pont{\matriz}{\espaco}{\B} + \pont{\matriz}{\B}{\A}$\,,
\item [(\textit{iii})] $\min \{ \pont{\matriz}{\A}{\C},
  \pont{\matriz}{\A}{\espaco} + \pont{\matriz}{\espaco}{\C} \} \le
  \pont{\matriz}{\A}{\B} + \pont{\matriz}{\B}{\C}$\,, and
\item [(\textit{iv})] $\max \{ \pont{\matriz}{\A}{\espaco},
  \pont{\matriz}{\espaco}{\A} \} \le \pont{\matriz}{\B}{\espaco} +
  \pont{\matriz}{\espaco}{\B}$\,,
\end{enumerate}
for each $\A, \B, \C \in \alphabet$.
\end{lemma}
\begin{proof}
Suppose that $\distanciaN{\matriz}(s, t) \le
\distanciaN{\matriz}(s, u) + \distanciaN{\matriz}(u, t)$ for all $s,
t, u \in \alphabet^{*}$. It follows from
Propositions~\ref{AuxdistanciaNTriangular-1},
\ref{AuxdistanciaNTriangular}, and~\ref{AuxdistanciaNTriangular-2}
that conditions (\textit{i}) to (\textit{iv})
are true.

Conversely, suppose that conditions (\textit{i}) to
(\textit{iv}) are true. Let $s, t, u \in \alphabet^{*}$ and $A, B$ be
N-optimal alignments of $s, u$ and $u, t$, respectively. It follows
from Proposition~\ref{desigTriangularGeral} that there exists an
alignment $C$ of $s, t$ and an integer $k \ge 0$ such that
\[
\custoA{\matriz}[A] + \custoA{\matriz}[B] \ge \custoA{\matriz}[C] + k
\Maior\,, \quad \tamanho{A} \le \tamanho{C} + k\,, \quad \text{and} \quad
\tamanho{B} \le \tamanho{C} + k\,.
\]
As a consequence of $\distanciaN{\matriz} \in \prametrica$, we have
$\custoA{\matriz}[A], \custoA{\matriz}[B] \ge 0$. It follows that
\[
\distanciaN{\matriz} (s, u) + \distanciaN{\matriz}(u,t) =
\frac{\custoA{\matriz}[A]}{\tamanho{A}} +
\frac{\custoA{\matriz}[B]}{\tamanho{B}} \ge \frac{\custoA{\matriz}[C]
  + k \Maior}{\tamanho{C} + k}\,.
\]
Since $\abs{C} >0$ and $k \ge 0$, we have from Fact~\ref{fato1} that
\begin{align*}
    \frac{\custoA{\matriz}[C]
  + k \Maior}{\tamanho{C} + k}  & \ge \min \left\{ \mathcal{Q}, \frac{\custoA{\matriz}[C]}{\tamanho{C}} = \custoN{\gamma}[C]
  \right\}\,,
\end{align*}
$\mathcal{Q} \ge \distanciaN{\gamma}(s, t)$ by Proposition~\ref{propAux2},
and $\custoN{\gamma}[C] \ge \distanciaN{\gamma}(s, t)$. Hence, 
\[
\distanciaN{\matriz} (s, t) \le \distanciaN{\matriz}(s,u) +
\distanciaN{\gamma}(u, t)\,.
\]
\end{proof}

Finally, using the results presented in this section, we can establish the following proof of the highlighted theorem.

\begin{proof} (of Theorem~\ref{theo:norm})

Suppose that $\distanciaN{\matriz} \in \metrica$. Consequently, for each $s, t, u \in \alphabet^{*}$, we have that $\distanciaN{\matriz}(s, s) = 0$,
$\distanciaN{\matriz}(s, t) > 0$ if $s \not= t$, $\distanciaN{\matriz}(s, t) =
\distanciaN{\matriz}(t, s)$, and $\distanciaN{\matriz}(s, u) \le
\distanciaN{\matriz}(s, t) + \distanciaN{\matriz}(t, u)$.

Let $\A, \B, \C \in \alphabet$, $\A \neq \B \neq \C$. Since
$\distanciaN{\matriz}(s, s) = 0$ and $\distanciaN{\matriz}(s, t) > 0$,
we have that $\distanciaN{\matriz} \in \prametrica$. Since
$\distanciaN{\matriz}(s, t) > 0$
if $s \not= t$ and $\distanciaN{\matriz} \in
\prametrica$, from Lemma~\ref{lema-maiorQueZero}, we have that
$\pont{\matriz}{\A}{\espaco}, \pont{\matriz}{\espaco}{\A},
\pont{\matriz}{\A}{\B} > 0$. Since $\distanciaN{\matriz} \in
\prametrica$, it follows from Corollary~\ref{corolaryPrametrica} that
$\pont{\matriz}{\A}{\A} = 0$ or $\pont{\matriz}{\A}{\espaco} +
\pont{\matriz}{\espaco}{\A} = 0$ and, since
$\pont{\matriz}{\A}{\espaco}, \pont{\matriz}{\espaco}{\A} > 0$, we
have that $\pont{\matriz}{\A}{\A} = 0$. Since $\distanciaN{\matriz}(s,
t) = \distanciaN{\matriz}(t, s)$ for each $s, t \in \alphabet^{*}$ and
$\distanciaN{\matriz} \in \prametrica$, from
Lemma~\ref{distanciaNSimetrica} we have that
$\pont{\matriz}{\A}{\espaco} = \pont{\matriz}{\espaco}{\A}$, and if
$\pont{\matriz}{\A}{\B} < \pont{\matriz}{\A}{\espaco}+
\pont{\matriz}{\espaco}{\B}$, then $\pont{\matriz}{\A}{\B} =
\pont{\matriz}{\B}{\A}$. Since $\distanciaN{\matriz}(s, u) \le
\distanciaN{\matriz}(s, t) + \distanciaN{\matriz}(t, u)$ and
$\distanciaN{\matriz} \in \prametrica$, from
Lemma~\ref{distanciaNTriangular} we have that
$\pont{\matriz}{\A}{\espaco} \le \pont{\matriz}{\A}{\B} +
\pont{\matriz}{\B}{\espaco}$ and $\min \{\pont{\matriz}{\A}{\C},
\pont{\matriz}{\A}{\espaco} + \pont{\matriz}{\espaco}{\C} \} \le
\pont{\matriz}{\A}{\B} + \pont{\matriz}{\B}{\C}$.

From the observations above, we have that if $\matriz \in \metricaN$,
then $\matriz \in \metricaA$. Besides that, since
$\pont{\matriz}{\A}{\espaco} = \pont{\matriz}{\espaco}{\A}$,
$\pont{\matriz}{\B}{\espaco} = \pont{\matriz}{\espaco}{\B}$ and $\max
\{ \pont{\matriz}{\A}{\espaco}, \pont{\matriz}{\espaco}{\A} \} \le
\pont{\matriz}{\B}{\espaco} + \pont{\matriz}{\espaco}{\B}$ from
Lemma~\ref{distanciaNTriangular}, we have that
\[
\pont{\matriz}{\A}{\espaco} = \max \{ \pont{\matriz}{\A}{\espaco},
\pont{\matriz}{\espaco}{\A} \} \le \pont{\matriz}{\B}{\espaco} +
\pont{\matriz}{\espaco}{\B} = \pont{\matriz}{\B}{\espaco}
+\pont{\matriz}{\B}{\espaco} = 2\,\pont{\matriz}{\B}{\espaco}\,.
\]

Conversely, suppose that $\matriz \in \metricaN$. By the definition of
$\metricaN$, we have that $\matriz \in \metricaA$ and
$\pont{\matriz}{\A}{\espaco} \le 2 \, \pont{\matriz}{\B}{\espaco}$
for each $\A, \B \in \alphabet$. Since $\matriz \in \metricaA$, we
have that $\distanciaA{\matriz} \in \metrica \subseteq \prametrica$
and $\pont{\matriz}{\A}{\espaco} = \pont{\matriz}{\espaco}{\A} > 0$;
$\pont{\matriz}{\A}{\B} > 0$ if $\A \not= \B$, and
$\pont{\matriz}{\A}{\B} = 0$ if $\A = \B$; if $\pont{\matriz}{\A}{\B}
< \pont{\matriz}{\A}{\espaco} + \pont {\matriz}{\espaco}{\B}$, then
$\pont{\matriz}{\A}{\B} = \pont{\matriz}{\B}{\A}$;
$\pont{\matriz}{\A}{\espaco} \le \pont{\matriz}{\A}{\B} +
\pont{\matriz}{\B}{\espaco}$; $\min \{ \pont{\matriz}{\A}{\C},
\pont{\matriz}{\A}{\espaco} + \pont {\matriz}{\espaco}{\C} \} \le
\pont{\matriz}{\A}{\B} + \pont{\matriz}{\B}{\C}$ for each $\A, \B, \C
\in \alphabet$.  In order to prove that $\distanciaN{\matriz} \in
\metrica$, we show, for each $s, t, u \in \alphabet^{*}$, that
$\distanciaN{\matriz}(s, s) = 0$, $\distanciaN{\matriz}(s, t) > 0$ for
$s \not= t$, $\distanciaN{\matriz}(s, t) =\distanciaN{\matriz}(t, s)$,
and $\distanciaN{\matriz}(s, u) \le \distanciaN{\matriz}(s, t) +
\distanciaN{\matriz}(t, u)$.

Since $\distanciaA{\matriz} \in \prametrica$, we have from
Lemma~\ref{lemaPrametricaN} that $\distanciaN{\matriz} \in
\prametrica$ and, thus, $\distanciaN{\matriz}(s, s) = 0$. Since
$\distanciaN{\matriz} \in \prametrica$, $\pont{\matriz}{\A}{\espaco} =
\pont{\matriz}{\espaco}{\A} > 0$, and $\pont{\matriz}{\A}{\B} > 0$ if
$\A \not= \B$, consequently from Lemma~\ref{lema-maiorQueZero} we have
that $\distanciaN{\matriz}(s, t) > 0$ for $s \not= t$. Since
$\distanciaN{\matriz} \in \prametrica$, $\pont{\matriz}{\A}{\espaco} =
\pont{\matriz}{\espaco}{\A}$, and 
$\pont{\matriz}{\A}{\B} = \pont{\matriz}{\B}{\A}$ if $\pont{\matriz}{\A}{\B} <
\pont{\matriz}{\A}{\espaco} + \pont {\matriz}{\espaco}{\B}$, we have from
Lemma~\ref{distanciaNSimetrica} that $\distanciaN{\matriz}(s,
t) =\distanciaN{\matriz}(t, s)$.

Since $\pont{\matriz}{\B}{\A} < \pont{\matriz}{\B}{\espaco} +
\pont{\matriz}{\espaco}{\A}$, we have that
$\pont{\matriz}{\B}{\A} = \pont{\matriz}{\A}{\B}$ whereas $\matriz \in \metricaA$.
Since $\pont{\matriz}{\A}{\espaco} = \pont{\matriz}{\espaco}{\A}$,
$\pont{\matriz}{\B}{\espaco} = \pont{\matriz}{\espaco}{\B}$, and
$\pont{\matriz}{\A}{\espaco} \le \pont{\matriz}{\A}{\B} +
\pont{\matriz}{\B}{\espaco}$, we have that
\[
\pont{\matriz}{\espaco}{\A} = \pont{\matriz}{\A}{\espaco} \le 
\pont{\matriz}{\A}{\B} +
\pont{\matriz}{\B}{\espaco} = \pont{\matriz}{\espaco}{\B} + 
\pont{\matriz}{\B}{\A}\,.
\]
Since $\pont{\matriz}{\A}{\espaco} = \pont{\matriz}{\espaco}{\A}$ and
$\pont{\matriz}{\B}{\espaco} = \pont{\matriz}{\espaco}{\B}$, it follows
from hypothesis that
\[
\max \{ \pont{\matriz}{\A}{\espaco}, \pont{\matriz}{\espaco}{\A} \} =
\pont{\matriz}{\A}{\espaco} \le 2\,\pont{\matriz}{\B}{\espaco} =
\pont{\matriz}{\B}{\espaco} + \pont{\matriz}{\espaco}{\B}\,.
\]

Since $\pont{\matriz}{\A}{\espaco} \le \pont{\matriz}{\A}{\B} +
\pont{\matriz}{\B}{\espaco}$, $\pont{\matriz}{\espaco}{\A} \le
\pont{\matriz}{\espaco}{\B} + \pont{\matriz}{\B}{\A}$, $\min \{
\pont{\matriz}{\A}{\C}, \pont{\matriz}{\A}{\espaco} +
\pont{\matriz}{\espaco}{\C} \} \le \pont{\matriz}{\A}{\B} +
\pont{\matriz}{\B}{\C},$ and $\max \{ \pont{\matriz}{\A}{\espaco},
\pont{\matriz}{\espaco}{\A} \} \le \pont{\matriz}{\B}{\espaco} +
\pont{\matriz}{\espaco}{\B}$, we have that $\distanciaN{\matriz}(s, u)
\le \distanciaN{\matriz}(s, t) + \distanciaN{\matriz}(t, u)$ from
Lemma~\ref{distanciaNTriangular}.
\end{proof}

Table~\ref{tabela3} summarizes properties of scoring matrices $\gamma$ for metric space and some other generalized metric space. 

\begin{table}[htpb]
\begin{minipage}{\textwidth}
\begin{center}
\begin{tabular}{clcccccc}
  & & $\prametrica$ & $\semimetrica$ & $\hemimetrica$ &
  $\pseudometrica$ & $\quasimetrica$ & $\metrica$ \\ \hline & & \\
  (a) & $\distanciaA{\gamma}$ is a premetric & \yes & \yes & \yes & \yes
  & \yes & \yes \\ & & \\
  (b) & $\pont{\matriz}{\A}{\espaco}, \pont{\matriz}{\espaco}{\A} > 0$
  and $\pont{\matriz}{\A}{\B} > 0$ if $\A \not= \B$ & & \yes & & &\yes
  & \yes \\& & \\
  (c) & $\pont{\matriz}{\A}{\espaco} = \pont{\matriz}{\espaco}{\A}$ &
  & \yes & & \yes & & \yes \\ & & \\
  (d) &
  \begin{tabular}{l}
    if $\pont{\matriz}{\A}{\B} < \pont{\matriz}{\A}{\espaco}
    + \pont{\matriz}{\espaco}{\B}$ then \\
    $\pont{\matriz}{\A}{\B} = \pont{\matriz}{\B}{\A}$
  \end{tabular}  
  & & \yes & & \yes & & \yes \\ & & \\
  (e) & $\pont{\matriz}{\A}{\espaco} \le \pont{\matriz}{\A}{\B} +
  \pont{\matriz}{\B}{\espaco}$ & & & \yes & \yes &\yes & \yes \\ & &
  \\
  (f) & $\pont{\matriz}{\espaco}{\A} \le \pont{\matriz}{\espaco}{\B} +
  \pont{\matriz}{\B}{\A}$ & & & \yes & \yes &\yes & \yes\\ & & \\
  (g) & 
  $\min \Big\{ \begin{array}{l}
    \pont{\matriz}{\A}{\C}, \\
    \pont{\matriz}{\A}{\espaco} + \pont{\matriz}{\espaco}{\C} 
  \end{array} \Big\} 
  \le \pont{\matriz}{\A}{\B} + \pont{\matriz}{\B}{\C}$
  &  &  & \yes & \yes &\yes & \yes\\ & & \\
  (h) &   $\max \{ \pont{\gamma}{\A}{\espaco}, \pont{\gamma}{\espaco}{\A} \}
  \le \pont{\matriz}{\B}{\espaco} + \pont{\matriz}{\espaco}{\B}$ & & & \yes & \yes &\yes & \yes
\end{tabular}
\end{center}
\end{minipage}
\caption{Necessary and sufficient conditions for scoring matrix
  $\matriz$ to induce $\distanciaN{\matriz}\text{-}p$ on sequences when $\distanciaN{\matriz} \in 
  \prametrica$.
  As in Table~\ref{tabela2},  
  the properties are also used to define metric ($\metrica$) and 
  generalized metric spaces
  such as \emph{premetric} ($\prametrica$),
  \emph{semimetric} ($\semimetrica$), \emph{hemimetric}
  ($\hemimetrica$), \emph{pseudometric} ($\pseudometrica$) and 
  \emph{quasimetric} ($\quasimetrica$).
  Results are obtained using
  definitions presented in Section~\ref{sec:preliminares} and
  lemmas in Section~\ref{sec:normalizado}.} \label{tabela3}
\end{table}

\section{Extended alignment of two sequences}\label{sec:estendido}

We describe in this section the classes of scoring matrices that induce $\distanciaE{\matriz}$-$p$ on sequences for each axiom $p$ of a metric. Lemmas~\ref{lema-ZeroE}--\ref{desigualdadeE} establish these properties, which are summarized in Table~\ref{tabela4}, and allow us to characterize matrices that induces each of the more general metric functions described in Section~\ref{sec:preliminares}. Lastly, we present the following important result previously stated in Section~\ref{sec:preliminares}:

\begin{theorem} \label{theo:extend} \rm
$\distanciaE{\matriz} \in \metrica$ if and only if $\matriz \in \metricaE$.
\end{theorem}

To prove this result, we proceed as in the previous sections and present some intermediary results as follows.

\begin{fact}\label{fato-permuta}\rm
Each weighted directed multigraph obtained by arcs that represent edit operations that transform a sequence into itself is Eulerian.
\end{fact}

\begin{lemma}\label{lema-ZeroE}\rm
Let $s \in \alphabet^{*}$ and $\matriz$ a scoring matrix. Then,
$\distanciaE{\matriz}(s, s) = 0$ if and only if $D(\matriz)$ has no
negative cycle.
\end{lemma}

\begin{proof}
Suppose that $\distanciaE{\matriz}(s, s) = 0$ and, by contradiction, 
$D(\matriz)$ has a cycle $W = x_0, \ldots, x_m$ with $x_m \not= \espaco$ and $\cost(W) = -X < 0$. 
Let $n$ be an integer such that 
$\pont{\matriz}{s(1)}{x_0} +
\pont{\matriz}{x_0}{s(1)} - nX < 0$.
Therefore, $A = [c, s(2), s(3), \ldots, s(\abs{s})]$,
where the column $c = s(1) x_0 (x_1, \ldots x_m)^n s(1)$, is also an extended alignment of 
$s, s$ and $\custoE{\gamma}[A] < 0 = \distanciaE{\matriz}(s, s)$,
which is a contradiction.
It follows that $D(\matriz)$ has no negative cycle.

On the other hand, suppose now that $D(\matriz)$ has no negative
cycle. Since $\custoE{\matriz}[s(1), s(2), \ldots, s(\abs{s})] = 0$, 
we have that $\distanciaE{\matriz}(s, s) \le 0$.

Let $A$ be an E-optimal alignment of $s, s$. 
From Fact~\ref{fato-permuta}, the multigraph $H$ obtained by considering
the edit operations that transform $s$ into itself is Eulerian, 
which implies that $H$ can be decomposed into cycles where each cycle is
also a cycle in $D(\gamma)$. 
Since $D(\matriz)$ has no negative cycle, we have that 
$\distanciaE{\matriz}(s,s) = \custoE{\matriz}[A] \ge 0$.

Consequently, since $\distanciaE{\matriz}(s, s) \le 0$ and
$\distanciaE{\matriz}(s, s) \ge 0$, we have that if $D(\matriz)$ has no negative cycle then $\distanciaE{\matriz}(s, s) = 0$.
\end{proof}

\begin{lemma}\label{lema-extMIZero}\rm
Let $s, t \in \alphabet^{*}$. Then, we have that
$\distanciaE{\matriz}(s, t) \ge 0$ if and only if
$\pont{\matriz}{\A}{\espaco}, \pont{\matriz}{\espaco}{\A},
\pont{\matriz}{\A}{\B} \ge 0$ for each $\A, \B \in \alphabet$.
\end{lemma}
\begin{proof}
Suppose that $\distanciaE{\matriz}(s, t) \ge 0$ for each $s, t \in
\alphabet^{*}$ and $\A, \B \in \alphabet$. Then,
$\pont{\matriz}{\A}{\espaco} = \custoE{\matriz} \alinhamento{
  \A, \espaco } \ge \distanciaE{\matriz}(\A, \seqVazia) \ge 0$.
Similarly, $\pont{\matriz}{\espaco}{\A}, \pont{\matriz}{\A}{\B} \ge
0$.

Conversely, suppose that $\pont{\matriz}{\A}{\espaco},
\pont{\matriz}{\espaco}{\A}, \pont{\matriz}{\A}{\B} \ge 0$ for each
$\A, \B \in \alphabet$. Thus, the sum of weights of any sequence of edit operations that transforms $s$ into $t$ is non negative. 
Therefore, $\distanciaE{\matriz}(s, t) \ge 0$.
\end{proof}

\begin{lemma}\label{lema-maiorZeroE}\rm
Let $s \not= t \in \alphabet^{*}$.  Then, $\distanciaE{\matriz}(s, t)
> 0$ if and only if
\begin{enumerate}
\item[(\textit{i})] $\pont{\matriz}{\A}{\A} \ge 0$\,, and
\item[(\textit{ii})] $\pont{\matriz}{\A}{\espaco},
  \pont{\matriz}{\espaco}{\A}, \pont{\matriz}{\A}{\B} > 0$\,,
\end{enumerate}
for each $\A \not= \B \in \alphabet$.
\end{lemma}
\begin{proof}
Suppose that $\distanciaE{\matriz} (s, t) > 0$. Then,
$\pont{\matriz}{\A}{\espaco} = \custoE{\matriz} \alinhamento{\A,
  \espaco} \ge \distanciaA{\matriz}(\A, \seqVazia) > 0$.
Similarly, $\pont{\matriz}{\espaco}{\A}, \pont{\matriz}{\A}{\B} > 0$.
Moreover, it follows from Lemma~\ref{lema-extMIZero} that
$\pont{\matriz}{\A}{\A} \ge 0$.

Conversely, suppose that (\textit{i}) and (\textit{ii}) are true. 
Since each edit operation that transforms $s$ into $t$ has
non negative weight and, because $s \not= t$, there exists at least
one edit operation with positive weight, we have that 
the sum of weights of any sequence of edit operations that transforms $s$ into $t$ is positive. 
Therefore, $\distanciaE{\matriz}(s, t) > 0$.
\end{proof}

\begin{lemma}\label{lema-simetriaE}\rm
Let $\matriz$ be a scoring matrix. Then, $\distanciaE{\matriz}(s, t)
= \distanciaE{\matriz}(t, s)$ for each $s, t \in \alphabet^{*}$ if
and only if $d_{\matriz}(\A, \B) = d_{\matriz}(\B, \A)$ and $d_{\matriz}(\A, \espaco) = d_{\matriz}(\espaco,
\A)$ for each $\A, \B \in \alphabet$.
\end{lemma}

\begin{proof}
Let $\A, \B \in \alphabet$. Suppose that $\distanciaE{\matriz}(s, t) =
\distanciaE{\matriz}(t, s)$ for each $s, t \in \alphabet^{*}$. Any
extended alignment of $\A, \seqVazia$ must have a single column.  Let
$\alinhamento{c}$ be an E-optimal alignment of $\A, \seqVazia$.  Then,
$c$ is a walk of minimum weight from $\A$ to $\espaco$ in
$D(\matriz)$. Hence, $d(\A, \espaco) =
\custoE{\matriz}\alinhamento{c} = \distanciaE{\matriz}(\A,
\seqVazia)$. Similarly, $\distanciaE{\matriz}(\seqVazia, \A) =
d(\espaco, \A)$. Therefore,
\[
d(\A, \espaco) = \distanciaE{\matriz}(\A, \seqVazia) =
\distanciaE{\matriz}(\seqVazia, \A) = d (\espaco, \A)\,.
\]

An E-optimal alignment of $\A, \B$ has either one or two columns. If
it has only one column $c$, we have that $d(\A, \B) = \distanciaE{\matriz}(\A, \B)$ as
stated in the previous paragraph. If the E-optimal alignment of $\A,
\B$ has two columns $c_{1}, c_{2}$, and $c$ is the walk of minimum
weight from $\A$ to $\B$, then we have, by the optimality of the
alignment, that
\[
\distanciaE{\matriz}(\A, \B) = \custoE{\matriz}\alinhamento{c_{1},
  c_{2}} \le \custoE{\matriz}\alinhamento{c} = d(\A, \B)\,.
\]
In the E-optimal alignment $\alinhamento{c_{1}, c_{2}}$, one of the
columns, say $c_{1}$, ends with $\espaco$ and $c_{2}$ begins with
$\espaco$.
Therefore, by concatenating the two columns, we obtain an extended
alignment $\alinhamento{c' = c_{1}(1)(=\A) \cdots c_{1}(m_{1}-1)
  \espaco c_{2}(2) \cdots c_{2}(m_{2}) (= \B)}$. This alignment has
only one column such that $\custoE{\matriz}\alinhamento{c_{1},
  c_{2}} = \custoE{\matriz}[c']$.  Since $c'$ is a walk from $\A$ to
$\B$, it follows that
\[
d(\A, \B) \le \custoE{\matriz}
\alinhamento{c'} = \custoE{\matriz}\alinhamento{c_{1}, c_{2}} = 
\distanciaE{\matriz}(\A, \B)\,.
\]
Thus, $d(\A, \B) = \distanciaE{\matriz}(\A, \B)$
also in this case. Similarly, $d(\B,
\A) = \distanciaE{\matriz}(\B, \A)$, which allows us to conclude,
since $\distanciaE{\matriz}(\A, \B) = \distanciaE{\matriz}(\B, \A)$,
that
\[
d(\A, \B) = \distanciaE{\matriz}(\A, \B) = 
\distanciaE{\matriz}(\A, \B) = \distanciaE{\matriz}(\B, \A)\,.
\]

Suppose now that $d(\A, \B) = d(\B, \A)$ for each $\A, \B \in
\alphabet_{\espaco}^{*}$. Let $\alinhamento{c_{1}, \ldots, c_{n}}$ be
an E-optimal alignment of $s, t$. Clearly,
$\alinhamento{c'_{1}, \ldots, c'_{n}}$ such that each $c'_{i}$ is a
walk of minimum weight from $c_{i}(m_{i})$ to $c_{i}(1)$ is an alignment
of $t, s$ and
\begin{align*}
\distanciaE{\matriz}(s, t) &= \sum_{i} \custoE{\matriz}[c_{i}] =
\sum_{i} d(c_{i}(1), c_{i}(m_{i})) \\
&= \sum_{i} d(c_{i}(m_{i}), c_{i}(1)) = \custoE{\matriz}
\alinhamento{c'_{1}, \ldots, c'_{n}}\\
&\le \distanciaE{\matriz}(t, s)\,. 
\end{align*}
Using similar reasoning, we have that $\distanciaE{\matriz}(t, s) \le
\distanciaE{\matriz}(s, t)$, which allows us to conclude that
$\distanciaE{\matriz}(s, t) = \distanciaE{\matriz}(t, s)$.
\end{proof}

\begin{lemma}\label{desigualdadeE}\rm
$\distanciaE{\matriz}(s, u) \le \distanciaE{\matriz}(s, t) +
\distanciaE{\matriz}(t, u)$ for any $s, t, u \in \alphabet^{*}$.
\end{lemma}
\begin{proof}
Let $\alinhamento{c^{st}_{1}, \ldots, c^{st}_{n_{st}}}$ and
$\alinhamento{c^{tu}_{1}, \ldots, c^{tu}_{n_{tu}}}$ be E-optimal
alignments of $s, t$ and $t, u$, respectively. Consider the sets of numbers $I = \{
i_{1}, \ldots, i_{\tamanho{t}} \}$ and $J = \{ j_{1}, \ldots,
j_{\tamanho{t}} \}$ such that $i_{1} < i_{2} < \ldots <
i_{\tamanho{t}}$, $j_{1} < j_{2} < \cdots < j_{\tamanho{t}}$, and $t(k)
= c^{st}_{i_{k}}(m_{i_{k}}) = c^{tu}_{j_{k}}(1)$.

Let $A$ be an alignment of $s, u$ whose columns are defined according
to the following rules: the column $c^{st}_{k}$ is a column of
$A$ for each $k \not\in I$; $c^{tu}_{k}$ is
a column of $A$ for each $k \not\in J$; and we define the column
\[
c^{st}_{i_{k}}(1) \ c^{st}_{i_{k}}(2)\  \cdots\ c^{st}_{i_{k}}(m_{i_{k}} - 1)
 \ t(k)\
c^{tu}_{j_{k}}(2) \ c^{tu}_{j_{k}}(3)\ \cdots\ c^{tu}_{j_{k}}(m_{j_{k}})\,,
\]
if $c^{st}_{i_k}(1) \not= \espaco$ or 
$c^{tu}_{j_{k}}(m_{j_{k}}) \not= \espaco$ 
for each $k = 1, \ldots, \tamanho{t}$.

Therefore, 
\begin{align*}
\distanciaE{\matriz}(s, u) &\le \custoE{\matriz}[A] \le
\custoE{\matriz}\alinhamento{c^{st}_{1}, \ldots, c^{st}_{n_{st}}} +
\custoE{\matriz}\alinhamento{c^{tu}_{1}, \ldots, c^{tu}_{n_{tu}}}\\
&= \distanciaE{\matriz}(s, t) + \distanciaE{\matriz}(t, u)\,. 
\end{align*}
\end{proof}

We are now able to present the proof of the main result of this section.

\begin{proof} (of Theorem~\ref{theo:extend})

Suppose first that $\distanciaE{\matriz} \in \metrica$. Then, for any
$s, t \in \alphabet^{*}, s \not= t$, we have that
$\distanciaE{\matriz}(s, s) = 0$, $\distanciaE{\matriz}(s, t) > 0$, and
$\distanciaE{\matriz}(s,t) = \distanciaE{\matriz}(t, s)$. 
From
Lemma~\ref{lema-maiorZeroE}, we have that $\pont{\matriz}{\A}{\A} \ge 0$ and $\pont{\matriz}{\A}{\B}, \pont{\matriz}{\A}{\espaco},
\pont{\matriz}{\espaco}{\A} > 0$ for each $\A \not= \B \in
\alphabet$ since $\distanciaE{\matriz}(s, t) > 0$. From Lemma~\ref{lema-simetriaE}, we have 
$d(\A, \B) = d(\B, \A)$ and $d(\A, \espaco) = d(\espaco, \A)$ for each
$\A, \B \in \alphabet$ since
$\distanciaE{\matriz}(s,t) = \distanciaE{\matriz}(t, s)$. It follows that
$\matriz \in \metricaE$.

Conversely, suppose that $\matriz \in \metricaE$. Then,
$\pont{\matriz}{\A}{\A} \ge 0$, $\pont{\matriz}{\A}{\B},
\pont{\matriz}{\A}{\espaco}, \pont{\matriz}{\espaco}{\A} > 0$, $d(\A,
\B) = d(\B, \A)$, and $d(\A, \espaco) = d(\espaco, \A)$ for each $\A
\not= \B$, $\A, \B \in \alphabet$. Since $\pont{\matriz}{\A}{\A} \ge
0$, $\pont{\matriz}{\A}{\B}, \pont{\matriz}{\A}{\espaco},
\pont{\matriz}{\espaco}{\A} > 0$ for each $\A \not= \B$, $\A, \B \in
\alphabet$, we have that $D(\matriz)$ has no negative cycle,
which implies, from Lemma~\ref{lema-ZeroE}, that
$\distanciaE{\matriz}(s, s) = 0$ for each $s \in \alphabet^{*}$.
Since $\pont{\matriz}{\A}{\A} \ge 0$, $\pont{\matriz}{\A}{\B},
\pont{\matriz}{\A}{\espaco}, \pont{\matriz}{\espaco}{\A} > 0$ we have,
from Lemma~\ref{lema-maiorZeroE}, that $\distanciaE{\matriz}(s, t) >
0$ for each $s, t \in \alphabet^{*}$, $s \not= t$. Since $d(\A, \B) =
d(\B, \A)$ and $d(\A, \espaco) = d(\espaco, \A)$ for each $\A \not=
\B$, $\A, \B \in \alphabet$, we have from Lemma~\ref{lema-simetriaE}
that $\distanciaE{\matriz}(s,t) = \distanciaE{\matriz}(t, s)$ for each
$s, t \in \alphabet^{*}$.  From Lemma~\ref{desigualdadeE}, we have
that $\distanciaE{\matriz}(s, u) \le \distanciaE{\matriz}(s, t) +
\distanciaE{\matriz}(t, u)$.  Therefore, $\distanciaE{\matriz} \in \metrica$.
\end{proof}

\begin{table}[htpb]
\begin{minipage}{\textwidth}
\begin{center}
\begin{tabular}{clcccccc}
  & & $\prametrica$ & $\semimetrica$ & $\hemimetrica$ &
  $\pseudometrica$ & $\quasimetrica$ & $\metrica$ \\ \hline & & \\
  (a) & $D(\matriz)$ has no negative cycle & \yes & \yes & \yes & \yes
  &\yes & \yes \\ & & \\
  (b) & $\pont{\matriz}{\A}{\espaco}, \pont{\matriz}{\espaco}{\B},
  \pont{\matriz}{\A}{\B} \ge 0$ & \yes & \yes & \yes & \yes &\yes &
  \yes \\ & & \\
  (c) & $\pont{\matriz}{\A}{\A} \ge 0$& & \yes & & &\yes & \yes \\ & &
  \\
  (d) & $\pont{\matriz}{\A}{\espaco}, \pont{\matriz}{\espaco}{\A} > 0$
  and $\pont{\matriz}{\A}{\B} > 0$ if $\A \not= \B$ & & \yes & & &\yes
  & \yes \\& & \\
  (e) & $d_{\matriz}( \A , \B) = d_{\matriz} (\B, \A)$ for each $\A, \B \in \Sigma_{\espaco}$ &
  & \yes & & \yes & & \yes \\ & & \\
\end{tabular}
\end{center}
\end{minipage}
\caption{Necessary and sufficient conditions for scoring matrix
  $\matriz$ to induce $\distanciaE{\gamma}\text{-}p$ on sequences where $p$ is each axiom of a metric. These properties are used to define metric spaces ($\metrica$) and generalizes metric spaces such as \emph{premetric} ($\prametrica$), \emph{semimetric} ($\semimetrica$), \emph{hemimetric}
  ($\hemimetrica$), \emph{pseudometric} ($\pseudometrica$), and
  \emph{quasimetric} ($\quasimetrica$). Results are obtained using
  definitions presented in Section~\ref{sec:preliminares} and lemmas in this section.} \label{tabela4}
\end{table}

\section{Conclusion}\label{sec:conclusion}

In this work we established necessary and sufficient conditions for
scoring matrices to induce each one of the properties of a metric in
weighted edit distances. For a subset of scoring matrices that induce
normalized edit distances, we also characterized the generalized metric spaces 
premetric, semimetric, hemimetric, pseudometric, quasimetric, and metric.
Moreover, we defined an
extended alignment distance, which takes into account a set of edit
operations that transforms one sequence into another, regardless the
existence of a corresponding standard alignment to represent it, describing a
criterion to find a sequence of edit operations whose weight is 
minimum. Similarly to the weighted edit distance, we determined the 
class of scoring matrices that induce extended alignment distances for 
each of the properties of a metric.

\bibliographystyle{elsarticle-harv} 
\bibliography{references}

\begin{thebibliography}{25}
\expandafter\ifx\csname natexlab\endcsname\relax\def\natexlab#1{#1}\fi
\providecommand{\url}[1]{\texttt{#1}}
\providecommand{\href}[2]{#2}
\providecommand{\path}[1]{#1}
\providecommand{\DOIprefix}{doi:}
\providecommand{\ArXivprefix}{arXiv:}
\providecommand{\URLprefix}{URL: }
\providecommand{\Pubmedprefix}{pmid:}
\providecommand{\doi}[1]{\href{http://dx.doi.org/#1}{\path{#1}}}
\providecommand{\Pubmed}[1]{\href{pmid:#1}{\path{#1}}}
\providecommand{\bibinfo}[2]{#2}
\ifx\xfnm\relax \def\xfnm[#1]{\unskip,\space#1}\fi
\bibitem[{Altschul et~al.(1990)Altschul, Gish, Miller, Myers and
  Lipman}]{Altschul.et.al.1990}
\bibinfo{author}{Altschul, S.F.}, \bibinfo{author}{Gish, W.},
  \bibinfo{author}{Miller, W.}, \bibinfo{author}{Myers, E.W.},
  \bibinfo{author}{Lipman, D.J.}, \bibinfo{year}{1990}.
\newblock \bibinfo{title}{Basic local alignment search tool}.
\newblock \bibinfo{journal}{Journal of Molecular Biology}
  \bibinfo{volume}{215}, \bibinfo{pages}{403--410}.
\bibitem[{Araujo et~al.(2021)Araujo, Rozante, Rubert and Martinez}]{ARRM2021}
\bibinfo{author}{Araujo, E.}, \bibinfo{author}{Rozante, L.C.},
  \bibinfo{author}{Rubert, D.P.}, \bibinfo{author}{Martinez, F.V.},
  \bibinfo{year}{2021}.
\newblock \bibinfo{title}{{Algorithms for Normalized Multiple Sequence
  Alignments}}, in: \bibinfo{editor}{Ahn, H.K.}, \bibinfo{editor}{Sadakane, K.}
  (Eds.), \bibinfo{booktitle}{Proceedings of the 32nd International Symposium
  on Algorithms and Computation (ISAAC 2021)}, \bibinfo{publisher}{Schloss
  Dagstuhl -- Leibniz-Zentrum f{\"u}r Informatik}, \bibinfo{address}{Dagstuhl,
  Germany}. pp. \bibinfo{pages}{40:1--40:16}.
\bibitem[{Araujo and Soares(2006)}]{AraujoS2006}
\bibinfo{author}{Araujo, E.}, \bibinfo{author}{Soares, J.},
  \bibinfo{year}{2006}.
\newblock \bibinfo{title}{Scoring matrices that induce metrics on sequences},
  in: \bibinfo{editor}{Correa, J.R.}, \bibinfo{editor}{Hevia, A.},
  \bibinfo{editor}{Kiwi, M.A.} (Eds.), \bibinfo{booktitle}{Proceedings of the
  Latin American Symposium on Theoretical Informatics (LATIN 2006)},
  \bibinfo{publisher}{Springer}. pp. \bibinfo{pages}{68--79}.
\bibitem[{Arkhangel'ski{\v \i} and Pontryagin(1990)}]{Topologia90}
\bibinfo{author}{Arkhangel'ski{\v \i}, A.V.}, \bibinfo{author}{Pontryagin,
  L.S.}, \bibinfo{year}{1990}.
\newblock \bibinfo{title}{General Topology {I}: Basic Concepts and
  Constructions, Dimension Theory}.
\newblock \bibinfo{publisher}{Springer-Verlag}, \bibinfo{address}{Berlin}.
\bibitem[{Barton et~al.(2015)Barton, Flouri, Iliopoulos and
  Pissis}]{BARTON2015}
\bibinfo{author}{Barton, C.}, \bibinfo{author}{Flouri, T.},
  \bibinfo{author}{Iliopoulos, C.S.}, \bibinfo{author}{Pissis, S.P.},
  \bibinfo{year}{2015}.
\newblock \bibinfo{title}{{Global and local sequence alignment with a bounded
  number of gaps}}.
\newblock \bibinfo{journal}{Theoretical Computer Science}
  \bibinfo{volume}{582}, \bibinfo{pages}{1--16}.
\bibitem[{Chaurasiya et~al.(2016)Chaurasiya, Londhe and Ghosh}]{GOSH2016}
\bibinfo{author}{Chaurasiya, R.K.}, \bibinfo{author}{Londhe, N.D.},
  \bibinfo{author}{Ghosh, S.}, \bibinfo{year}{2016}.
\newblock \bibinfo{title}{{A novel weighted edit distance-based spelling
  correction approach for improving the reliability of Devanagari script-based
  P300 speller system}}.
\newblock \bibinfo{journal}{{IEEE Access}} \bibinfo{volume}{4},
  \bibinfo{pages}{8184--8198}.
\bibitem[{Chenna et~al.(2003)Chenna, Sugawara, Koike, Lopez, Gibson, Higgins
  and Thompson}]{Chenna.et.al.2003}
\bibinfo{author}{Chenna, R.}, \bibinfo{author}{Sugawara, H.},
  \bibinfo{author}{Koike, T.}, \bibinfo{author}{Lopez, R.},
  \bibinfo{author}{Gibson, T.J.}, \bibinfo{author}{Higgins, D.G.},
  \bibinfo{author}{Thompson, J.D.}, \bibinfo{year}{2003}.
\newblock \bibinfo{title}{Multiple sequence alignment with the clustal series
  of programs}.
\newblock \bibinfo{journal}{Nucleic Acids Research} \bibinfo{volume}{31},
  \bibinfo{pages}{3497--3500}.
\bibitem[{Fisman et~al.(2022)Fisman, Grogin, Margalit and Weiss}]{FISMAN2022}
\bibinfo{author}{Fisman, D.}, \bibinfo{author}{Grogin, J.},
  \bibinfo{author}{Margalit, O.}, \bibinfo{author}{Weiss, G.},
  \bibinfo{year}{2022}.
\newblock \bibinfo{title}{{The Normalized Edit Distance with Uniform Operation
  Costs is a Metric}}.
\newblock \bibinfo{journal}{{arXiv preprint arXiv:2201.06115}} .
\bibitem[{de~la Higuera and Mic{\'o}(2008)}]{MICO2008}
\bibinfo{author}{de~la Higuera, C.}, \bibinfo{author}{Mic{\'o}, L.},
  \bibinfo{year}{2008}.
\newblock \bibinfo{title}{{A Contextual Normalised Edit Distance}}, in:
  \bibinfo{booktitle}{{2008 IEEE 24th International Conference on Data
  Engineering Workshop}}, \bibinfo{organization}{IEEE}. pp.
  \bibinfo{pages}{354--361}.
\bibitem[{Karplus et~al.(1998)Karplus, Barrett and Hughey}]{KBH1998}
\bibinfo{author}{Karplus, K.}, \bibinfo{author}{Barrett, C.},
  \bibinfo{author}{Hughey, R.}, \bibinfo{year}{1998}.
\newblock \bibinfo{title}{{Hidden Markov models for detecting remote protein
  homologies}}.
\newblock \bibinfo{journal}{Bioinformatics} \bibinfo{volume}{14},
  \bibinfo{pages}{846--856}.
\bibitem[{Kim(1968)}]{pseudoQuasi1968}
\bibinfo{author}{Kim, Y.}, \bibinfo{year}{1968}.
\newblock \bibinfo{title}{Pseudo quasi metric spaces}.
\newblock \bibinfo{journal}{Proceedings of the Japan Academy, Series A,
  Mathematical Sciences} \bibinfo{volume}{44}, \bibinfo{pages}{1009--1012}.
\bibitem[{Levenshtein(1965)}]{Levenshtein1965}
\bibinfo{author}{Levenshtein, V.I.}, \bibinfo{year}{1965}.
\newblock \bibinfo{title}{Binary codes capable of correcting deletions,
  insertions and reversals}.
\newblock \bibinfo{journal}{Soviet Physics Doklady} \bibinfo{volume}{10},
  \bibinfo{pages}{707--710}.
\bibitem[{Lipman et~al.(1989)Lipman, Altschul and Kececioglu}]{LAK1989}
\bibinfo{author}{Lipman, D.J.}, \bibinfo{author}{Altschul, S.F.},
  \bibinfo{author}{Kececioglu, J.D.}, \bibinfo{year}{1989}.
\newblock \bibinfo{title}{A tool for multiple sequence alignment}.
\newblock \bibinfo{journal}{Proceedings of the National Academy of Sciences}
  \bibinfo{volume}{86}, \bibinfo{pages}{4412--4415}.
\bibitem[{Lipman and Pearson(1985)}]{LP1985}
\bibinfo{author}{Lipman, D.J.}, \bibinfo{author}{Pearson, W.R.},
  \bibinfo{year}{1985}.
\newblock \bibinfo{title}{Rapid and sensitive protein similarity searches}.
\newblock \bibinfo{journal}{Science} \bibinfo{volume}{227},
  \bibinfo{pages}{1435--1441}.
\bibitem[{Marzal and Vidal(1993)}]{MarzalV1993}
\bibinfo{author}{Marzal, A.}, \bibinfo{author}{Vidal, E.},
  \bibinfo{year}{1993}.
\newblock \bibinfo{title}{Computation of normalized edit distance and
  applications}.
\newblock \bibinfo{journal}{IEEE Transactions on Pattern Analysis and Machine
  Intelligence} \bibinfo{volume}{15}, \bibinfo{pages}{926--932}.
\bibitem[{Needleman and Wunsch(1970)}]{NeedlemanW1970}
\bibinfo{author}{Needleman, S.B.}, \bibinfo{author}{Wunsch, C.D.},
  \bibinfo{year}{1970}.
\newblock \bibinfo{title}{A general method applicable to the search for
  similarities in the amino acid sequence of two proteins.}
\newblock \bibinfo{journal}{Journal of Molecular Biology} \bibinfo{volume}{48},
  \bibinfo{pages}{443--453}.
\bibitem[{Notredame et~al.(2000)Notredame, Higgins and Heringa}]{NHH2000}
\bibinfo{author}{Notredame, C.}, \bibinfo{author}{Higgins, D.G.},
  \bibinfo{author}{Heringa, J.}, \bibinfo{year}{2000}.
\newblock \bibinfo{title}{T-coffee: a novel method for fast and accurate
  multiple sequence alignment}.
\newblock \bibinfo{journal}{Journal of Molecular Biology}
  \bibinfo{volume}{302}, \bibinfo{pages}{205--217}.
\bibitem[{Pearson(2013)}]{Pea2013}
\bibinfo{author}{Pearson, W.R.}, \bibinfo{year}{2013}.
\newblock \bibinfo{title}{Selecting the right similarity-scoring matrix}.
\newblock \bibinfo{journal}{Current Protocols in Bioinformatics}
  \bibinfo{volume}{43}, \bibinfo{pages}{3.5.1--3.5.9.}
\bibitem[{Sellers(1974)}]{Sellers1974}
\bibinfo{author}{Sellers, P.H.}, \bibinfo{year}{1974}.
\newblock \bibinfo{title}{On the theory and computation of evolutionary
  distances}.
\newblock \bibinfo{journal}{SIAM Journal on Applied Mathematics}
  \bibinfo{volume}{26}, \bibinfo{pages}{787--793}.
\bibitem[{Smith and Waterman(1981)}]{SW1981}
\bibinfo{author}{Smith, T.F.}, \bibinfo{author}{Waterman, M.S.},
  \bibinfo{year}{1981}.
\newblock \bibinfo{title}{Identification of common molecular subsequences}.
\newblock \bibinfo{journal}{Journal of Moelcular Biology}
  \bibinfo{volume}{147}, \bibinfo{pages}{195--197}.
\bibitem[{Steen and Seebach(1978)}]{pseudo-1970}
\bibinfo{author}{Steen, L.A.}, \bibinfo{author}{Seebach, J.A.},
  \bibinfo{year}{1978}.
\newblock \bibinfo{title}{Counterexamples in Topology}.
\newblock \bibinfo{edition}{2nd} ed., \bibinfo{publisher}{Springer-Verlag}.
\bibitem[{Sun et~al.(2019)Sun, Ni, Chng, Liu, Luo, Ng, Hanu, Ding, Liu,
  Karatzas, Chan and Jin}]{SUN2019}
\bibinfo{author}{Sun, Y.}, \bibinfo{author}{Ni, Z.}, \bibinfo{author}{Chng,
  C.K.}, \bibinfo{author}{Liu, Y.}, \bibinfo{author}{Luo, C.},
  \bibinfo{author}{Ng, C.C.}, \bibinfo{author}{Hanu, J.},
  \bibinfo{author}{Ding, E.}, \bibinfo{author}{Liu, J.},
  \bibinfo{author}{Karatzas, D.}, \bibinfo{author}{Chan, C.S.},
  \bibinfo{author}{Jin, L.}, \bibinfo{year}{2019}.
\newblock \bibinfo{title}{{ICDAR 2019 competition on large-scale street view
  text with partial labeling-RRC-LSVT}}, in: \bibinfo{booktitle}{{2019
  International Conference on Document Analysis and Recognition (ICDAR)}},
  \bibinfo{organization}{{IEEE}}. pp. \bibinfo{pages}{1557--1562}.
\bibitem[{Wilson(1931a)}]{Wilson31-quasi}
\bibinfo{author}{Wilson, W.}, \bibinfo{year}{1931}a.
\newblock \bibinfo{title}{On quasi-metric spaces}.
\newblock \bibinfo{journal}{American Journal of Mathematics}
  \bibinfo{volume}{43}, \bibinfo{pages}{675--684}.
\bibitem[{Wilson(1931b)}]{Wilson31-semi}
\bibinfo{author}{Wilson, W.}, \bibinfo{year}{1931}b.
\newblock \bibinfo{title}{On semi-metric spaces}.
\newblock \bibinfo{journal}{American Journal of Mathematics}
  \bibinfo{volume}{53}, \bibinfo{pages}{361--373}.
\bibitem[{Yujian and Bo(2007)}]{YujianB2007}
\bibinfo{author}{Yujian, L.}, \bibinfo{author}{Bo, L.}, \bibinfo{year}{2007}.
\newblock \bibinfo{title}{{A Normalized Levenshtein Distance Metric}}.
\newblock \bibinfo{journal}{IEEE Transactions on Pattern Analysis and Machine
  Intelligence} \bibinfo{volume}{29}, \bibinfo{pages}{1091--1095}.

\end{thebibliography}






\end{document}